\documentclass[preprint2]{emulateapj}
\usepackage{natbib}
\usepackage{graphicx}
\usepackage{pdfpages}

\usepackage{CJK}

\newcommand{\hii}{\mbox{H{\sc ii}~}}
\newcommand{\hi}{\mbox{H{\sc i}~}}

\begin{document}

\title{The low-mass  population in the young cluster Stock 8: Stellar properties and Initial Mass Function}
\begin{CJK*}{UTF8}{gbsn}

%Gregory J. Herczeg (沈雷歌)

\end{CJK*}

\author{Jessy Jose  \altaffilmark{1},
  \begin{CJK*}{UTF8}{gbsn}
    Gregory J. Herczeg(沈雷歌)\altaffilmark{1}, Manash R. Samal\altaffilmark{2}, Qiliang Fang(方其亮)\altaffilmark{1}\end{CJK*} and Neelam Panwar\altaffilmark{3}  }
\altaffiltext{1}{Kavli Institute for Astronomy and Astrophysics, Peking University, Yi He Yuan Lu 5, Haidian Qu, Beijing 100871, China; jessyvjose1@gmail.com}
\altaffiltext{2}{Graduate Institute of Astronomy, National Central University 300, Jhongli City, Taoyuan County - 32001, Taiwan}
\altaffiltext{3}{Department of Physics \& Astrophysics, University of Delhi, Delhi - 110007, India}

\begin{abstract}

  The evolution of \hii regions/supershells can trigger a new generation of stars/clusters at their peripheries, with environmental conditions that may affect the initial mass function, disk evolution and star formation efficiency. In this paper we study the stellar content and star formation processes in the young  cluster Stock 8, which itself  is thought to be formed during the expansion of a supershell. We present deep optical photometry along with JHK and 3.6, 4.5 $\mu$m photometry from UKIDSS and  {\it Spitzer}-IRAC. We use multi-color criteria to identify the candidate young stellar objects  in the region. Using evolutionary models, we obtain a median log(age) of $\sim$6.5 ($\sim$3.0 Myr)  with an  observed age spread of $\sim$0.25 dex for the cluster. Monte Carlo simulations of the population of Stock 8, based on estimates for the photometric uncertainty, differential reddening, binarity, and variability, indicate that these uncertainties introduce an age spread of $\sim$0.15 dex. The  intrinsic age spread   in the cluster is $\sim$0.2 dex. The fraction of young stellar objects surrounded by disk is $\sim$35\%.  The K-band luminosity function of Stock 8 is  similar to that of the Trapezium cluster. The IMF of  Stock 8 has a Salpeter-like  slope at $>$ 0.5 $M_{\odot}$ and the IMF flattens and   peaks at $\sim$0.4 $M_{\odot}$, below which declines into the substellar regime. Although Stock 8 is surrounded  by several massive stars, there seems to be no severe environmental effect in the form of IMF  due to the proximity of massive stars around the cluster.

\end{abstract}
\keywords{ISM: individual objects (Stock 8) --stars: formation -- stars: pre-main sequence}

\section{Introduction}
\label{intro}

The initial mass function (IMF) is an important statistical tool to understand the  formation of  stars and brown dwarfs.
Whether the IMF is universal with a shape similar to that of the  Galactic field  with a general flattening below 1 M$_\odot$
or  is a function of star formation environment  is a  question yet to be answered \citep{bastian2010}. 
Numerical simulations suggest that the peak stellar  mass and the associated plateau of the IMF 
of a given region is related to the thermal Jeans mass at the onset of isothermal collapse (e.g., \citealt{bate2005}). 
The plateau properties would then vary with the environmental conditions,  for instance the dependence of 
the Jeans mass on the density and temperature in the molecular core. 
However, analysis including additional physics, in particular radiative
feedback, suggests that  the variations in the IMF should be modest (\citealt{bate2009,Krumholz2012}). 

Although there is evidence showing that the IMF appears to be constant across different star forming regions and 
even in the sub-stellar regime (\citealt{bastian2010,offner2014}), the topic is still under debate (e.g. \citealt{dib2014}).
 Evidence for IMF variations at high-masses has been presented for extreme regions such as the Galactic center (e.g. \citealt{lu2013}).
Similarly, high velocity dispersion galaxies have systematically bottom heavy IMF compared to Galactic field IMF and hence IMF may be a function of               
environmental conditions \citep{vandokkum2010}. The  IMF is well characterized in most of the nearby 
low mass  star forming regions such as Taurus or Ophiuchus (see review by \citealt{luhman2012}). However the conditions of
star formation vary considerably between the massive star forming regions,  which contain several tens to hundreds  of OB stars, and
low mass star forming regions. Radiation and stellar winds from O-type stars may have a profound effect on nearby star formation
(e.g., \citealt{deharveng2012,sicilia2014,panwar2014,ojha2011})
by regulating the gas dynamics, density, temperature and radiation field. As a result, the products  of  star formation, such as the IMF,
binary fraction and the evolution of protostellar disks,  are likely to vary from region to region. 

Whether the IMF is universal may be tested with its measurable properties, including the flattening and peak position of
IMF in massive star environments. Young clusters 
located in the vicinity of the large Galactic bubbles, where the local environment might have been  affected by  feedback 
from nearby massive stars, are ideal targets for studying  the environmental  influence on star formation process.
If these effects are significant, they may appear in the  young cluster Stock 8. 
Stock 8 is  located in the   Perseus arm and is part of the extended \hii region
Sh2-234 (IC417).  \citet{jose2008} (hereafter Paper 1) used optical $UBVI$ and 2MASS $JHK$ photometry 
to  derive the cluster  parameters such as the cluster center ($\alpha_{2000}$ = $05^{h}28^{m}07^{s}$;
$\delta_{2000}$ = $+34^{\circ}25^{\prime}42^{\prime\prime}$), distance ($\sim$ 2.05 kpc) and age ($\sim$ 1-5 Myr).
 Paper 1 shows that Stock 8 is surrounded by 
12 massive OB type stars and is probably part of a
large OB association. Based on the  strong ionized  gas emission at 1.4 GHz as well as  in $H\alpha$ in the cluster
vicinity, Paper 1 argues  that the cluster and its immediate neighbourhood are under the strong  influence  of the surrounding
massive stars. A recent photometric and spectroscopic survey of  $\sim$ 25 pc radius around the cluster by \citet{marco2016}
further reveals that the cluster is surrounded by at least 33 early type massive stars ($>$ $B2V$). 
These properties make the cluster a special target  to study the effect of massive stellar feedback from the early epoch of star formation 
to  the cluster evolution process as well as the form of IMF in the sub-stellar regime.

In Paper 1 the cluster member  list was limited by the shallow  observations in optical and $JHK$ bands. Hence the low-mass
population of Stock 8    largely remains  unexplored so far.  The low-mass stars constitute the  majority of the stellar population in
the Galaxy \citep{kroupa2002}.
Deep near-infrared observations are an ideal tool to uncover the low mass stellar content of 
the distant regions. In this study, we investigate the low mass young stellar content of the  cluster Stock 8 using deep, high resolution
optical observations ($V$ $\sim$ 24 mag) in combination with the UKIDSS $JHK$ and {\it Spitzer}-IRAC data sets.
The  main objectives  of this work are: 1) to identify and characterize the YSOs, 2) construct the  IMF of Stock 8 down
to  the low mass regime and to test whether there is any influence of massive stars in its vicinity to the form of IMF.  
The distance estimated for the cluster in Paper 1 (2.1 kpc) is adopted here and  is consistent with the distance of $\sim$ 1.8 -- 2.3 kpc
obtained by \citealt{pandey2013,kharchenko2013,hou2014,foster2015} towards this region.
The paper is organized as follows: in Section 2 we describe the various data sets 
used and generation of point source catalog. In Sections 3 and 4 we characterize the YSOs of the region and derive the various cluster 
properties. Section 5 describes the luminosity function and mass function analysis,  Section 6 discusses the large scale structure of
the complex and  Section 7 concludes the various results obtained.

\section{OBSERVATIONS AND POINT SOURCE CATALOGS}

One of the main goals of this study is to obtain the census of YSOs in the cluster. The identification and classification of
YSOs  in the region are mainly based on  $JHK$ and IRAC-3.6 and 4.5 $\mu$m photometry (see Section \ref{yso}). However,  the
4.5 $\mu$m data is limited to 3$^\prime$ radius of the cluster towards its north. Hence in this study we focus an area of 3$^\prime$ radius
around the cluster. In this section we describe the individual data sets used for this study. 

%%%%%%%%%%%%%%%%%%%%%%%%%%%%%%%%%%%%%%%%%%%%%%%%%%%%%%%%%%%%%%%%%%%%%%%%%%%%%%%%%%%%%%%%%%%%%%%%%%
\begin{figure*}[t]
\centering
\includegraphics[scale=0.4,trim=0 120 0  60,clip]{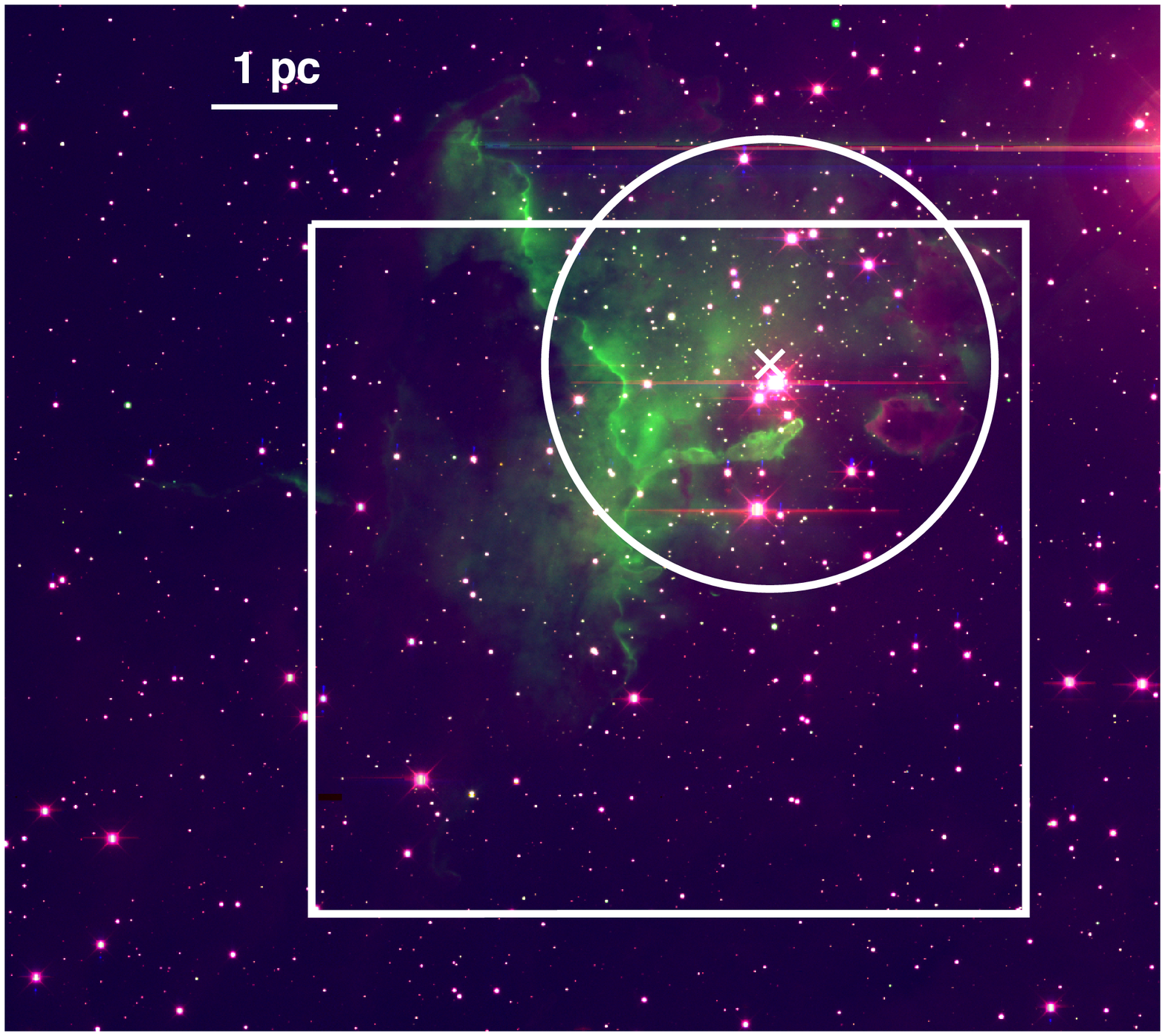}
\includegraphics[scale = 0.45, trim = 10 160 10 200, clip]{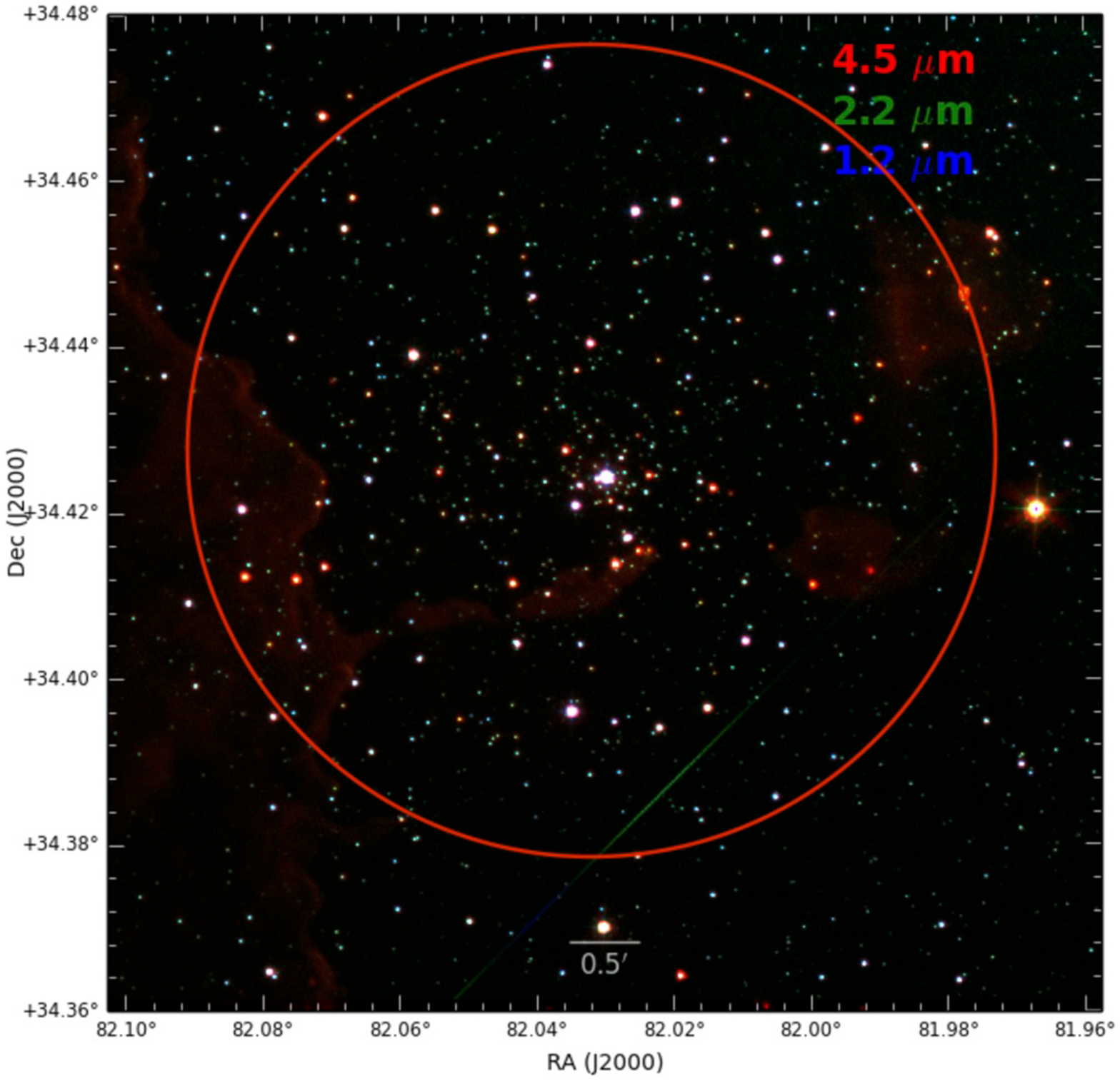}
\caption{ {\it left}: Color composite image of the cluster Stock 8 made using $B$ (blue), $H{\alpha}$ (green) and $I$-band
  images from the NOAO archive. The cluster center is represented by the cross mark, the circle represents the 3$^{\prime}$ radius
  around the center and  the box represents the area covered by the SNIFS observations in $V$ and $I$ bands. {\it right}:
  Color composite image made from 4.5 $\mu$m (IRAC: red), 2.2 $\mu$m (UKIDSS: green) and 1.2 $\mu$m (UKIDSS: blue)  bands and
the circle represents the  3$^{\prime}$ radius. 
  }
\label{area}
\end{figure*}
%%%%%%%%%%%%%%%%%%%%%%%%%%%%%%%%%%%%%%%%%%%%%%%%%%%%%%%%%%%%%%%%%%%%%%%%%%%%%%%%%%%%%%%%%%%%

\subsection{Optical Photometry}
\label{optical}

Optical imaging  of Stock 8 was obtained on 13 December 2014 with the SNIFS instrument on the UH 2.2 m 
telescope at Mauna Kea, Hawaii. The instrument covers an area $\sim$ 9$^{\prime}$.2 $\times$ 9$^{\prime}$.2 on the sky 
with a plate scale of 0$^{\prime\prime}$.273 per pixel.  Two exposures were obtained in $V$ and $I$ bands, with integration times of
150 and 100 seconds, respectively.  Average seeing  during observations
was $\sim$  0$^{\prime\prime}$.7.  Bias and flat frames taken on the same night were used to pre-process the images.  
The area coverage of this observation is shown by a  box in Fig. \ref{area}.

Images were reduced and analyzed  using IRAF software. The sources above 5$\sigma$ of the background
were detected using DAOFIND task in IRAF.  Bright, isolated  stars were selected across the field to construct 
a characteristic point-spread function (PSF) for the images.  The PSF photometry of all the sources were obtained 
using ALLSTAR task in IRAF.  The absolute photometry calibration in $V, I$ bands was obtained using the published photometry from Paper 1. 
We matched the bright isolated sources  in the SNIFS catalog with that of Paper 1 within a radius of 1$^{\prime\prime}$.0 and obtained the
color and  magnitude coefficients in $V$ and $I$ bands. These coefficients were applied to the entire SNIFS catalog  to obtain the
calibrated magnitudes. The saturated sources in 
the SNIFS photometry are replaced with the magnitudes given in Paper 1. We finally consider only those sources with uncertainty $<$0.2 mag
in $I$ band for further analysis. The number of sources detected within 3$^\prime$ radius of Stock 8 is given in  Table \ref{catalog}.
The detection limits are 24.2 and 22.8 mag in $V$ and $I$ bands, respectively (see Table \ref{catalog}), which
is $\sim$ 2 mag deeper than Paper 1. In terms of spatial resolution and sensitivity, this is the deepest optical photometric survey of
the region to date. 

\subsection{NIR photometry from UKIDSS}

NIR photometry was  obtained from the UKIDSS 6$^{th}$ archival data release (UKIDSSDR6plus) of the Galactic Plane Survey
(GPS; \citealt{lucas2008}). 
UKIDSS observations were performed using the UKIRT Wide Field Camera (WFCAM; \citealt{casali2007}). Typical 
UKIDSS PSFs have 0$^{\prime\prime}$.8 -- 1$^{\prime\prime}$ FWHM
with 0$^{\prime\prime}$.4 pixels.  2MASS photometry  is used to calibrate the  final fluxes.
Details about the data reduction and calibration procedures are given in 
\citet{dye2006} and \citet{hodgkin2009}, respectively. In order to select  reliable point sources from the catalog, we consider
only those sources with uncertainty $<$0.2 mag  in all three bands.  
The list was again visually checked for any spurious source detection and such sources were deleted from the list.
The magnitudes of the saturated bright sources were  retrieved from the 2MASS catalog. The final NIR catalog includes  
1256  sources common to  $J$, $H$, and $K$ bands within 3${^\prime}$ radius of the cluster and
the detection limits of individual bands are given in Table \ref{catalog}. UKIDSS photometry is $\sim$ 4 mag deeper than 2MASS photometry
in all three bands.

%%%%%%%%%%%%%%%%%%%%%%%%%%%%%%%%%%%%%%%%%%%%%%%%%%%%%%%%%%%%%%%%%%%%%%%%%%
\begin{center}
\begin{table}
\caption{Point source catalog summary within 3$^{\prime}$ radius of the cluster}
\label{catalog}
\centering
%\begin{tabular}{|p{.40in}|p{.45in}|p{.58in}|p{.28in}|}
\begin{tabular}{ccccc}
%\hline
\tableline\tableline
 
Band & number of    &  detection &completeness & completeness\\
  & sources &  limit (mag) & limit (mag) & limit (M$_\odot$) \\
  \tableline

$V$ & 848 & 24.2  & 23.0 & 0.3  \\
$I$ & 956 & 22.8  & -    & - \\
$J$ & 1256 & 20.0 & 19.0 & 0.08 \\
$H$ & 1259 & 19.3 & -    & - \\
$K$ & 1261 & 18.3 & 18.0 & 0.08 \\
ch1 &  865 & 17.2 & 15.0 & 0.4 \\
ch2 &  868 & 17.1 & 15.0 & 0.4\\

\tableline
%%\hline
\end{tabular}
%\end{tabular}
\end{table}
\end{center}
%%%%%%%%%%%%%%%%%%%%%%%%%%%%%%%%%%%%%%%%%%%%%%%%%%%%%%%%%%%%%%%%%%%%%%%%%%

\subsection{{\it Spitzer}-IRAC Photometry}
\label{spitzer}

The {\it Spitzer} data in the two bands of IRAC centred  at 3.6 and 4.5  $\mu$m (ch1 and ch2) was obtained from the
{\it Spitzer} archive\footnote  {http://archive. spitzer.caltech.edu/}. These observations were taken  on 2010 May 05
(Program ID: 61070 PI: Barbara Whitney) as part of the {\it Spitzer} warm mission, in high dynamic range (HDR) mode with
two dithers per map  position and two images each with integration time of 0.4s and 10.4s per dither. 

We obtained the cBCD ({\it corrected basic calibrated data}) 
images   from the  archive and the raw data was processed and calibrated with the IRAC pipeline. The final mosaic images
were created using the MOPEX pipeline  with an image scale of 1$^{\prime\prime}$.2 per pixel.
We used the DAOFIND task in IRAF to  create a preliminary source list with the condition 5$\sigma$ above  the background in ch2.  
The spurious sources were visually identified and deleted from the list. To extract the flux of the point sources, 
we performed  point response function  (PRF) fitting on IRAC images in multi frame mode, using the Astronomical Point
Source EXtraction (APEX) tool, developed by the {\it Spitzer}  Science Centre. Flux densities are converted in to magnitudes
using the  zero-points 280.9 and  179.7   Jys in the 3.6 and  4.5 $\mu$m bands, respectively, following the IRAC Data
Handbook\footnote{http://ssc.spitzer.caltech.edu/irac/iracinstrumenthandbook/}. The saturated bright sources in the long
integrated images were replaced by the sources from  short images.  The IRAC data of both the  bands were merged by matching
the coordinates using a radial matching tolerance of 1$^{\prime\prime}$.2. Only those sources with uncertainty $<$0.2 mag
are used for further analysis. The number of sources detected in each band and the detection limits  are given in Table \ref{catalog}.

\subsection{Final catalog}

We created the final photometry catalog by spatially matching and merging the detected sources in various bands. 
We obtained  the NIR detection for all the IRAC sources in both channels within a match radius of 1$^{\prime\prime}$.
A search for the optical counterparts of the NIR sources within a radius of  1$^{\prime\prime}$ identified 
838 and 946 sources in $V$ and $I$ bands, respectively. This shows that $\sim$ 99\% of our optical sources have NIR counterparts. 
Sample entries of photometric data of all the stars within 3$^\prime$ radius of Stock 8 is given in Table \ref{phottable}.
The full table is available in the electronic version of the journal.

%%%%%%%%%%%%%%%%%%%%%%%%%%%%%%%%%%%%%%%%%%%%%%%%%%%%%%%%%%%%%%%%%%%%%%%%%%

\begin{table*}
\centering
\caption{Photometric data$^a$ of all point sources  within 3$^\prime$ radius of Stock 8. }
\label{phottable}
\begin{center}
%\scriptsize
%\begin{tabular}{|p{.40in}p{.4in}p{.4in}p{.4in}p{.4in}p{.4in}p{.4in}p{.4in}p{.4in}p{.4in}p{.4in}|}
\begin{tabular}{cccccccccc}
 \hline

 $\alpha_{(2000)}$& $\delta_{(2000)}$ & $V$ & $I$ &$J$ & $H$ & $K$ & $[3.6]$ & $[4.5]$  & Class  \\
 deg        &  deg   & mag    & mag       & mag       & mag & mag & mag & mag &   \\
\hline

 82.0033 & +34.4227  &  22.10  &    19.02     &        16.99    &     16.15    &      15.71  &       15.26    &      14.90    & Class II \\      
 82.0527 & +34.4321  &  22.30  &    19.23     &        14.46    &     13.48    &      12.99  &       12.17    &      11.86    & Class II \\     
 82.0143 & +34.4228  &  17.72  &    15.73     &        13.77    &     12.86    &      12.20  &       11.21    &      10.79    & Class II  \\    
 81.9839 & +34.4582  &  19.42  &    17.81     &        16.62    &     16.10    &      15.88  &       15.54    &      15.15    & Class II  \\    
 82.0364 & +34.4367  &  19.74  &    17.36     &        15.48    &     14.62    &      14.20  &       13.65    &      13.36    & Class II  \\ 

\hline
\end{tabular}\\
$^a$ The complete table is available in electronic form.
\end{center}
\end{table*}

%%%%%%%%%%%%%%%%%%%%%%%%%%%%%%%%%%%%%%%%%%%%%%%%%%%%%%%%%%%%%

\subsection{Completeness limits of the data}
\label{completeness}
Artificial star tests are a standard procedure to assess the level of completeness and accuracy of a photometric analysis.
We performed the artificial star experiment  using the ADDSTAR tool in IRAF in $V$, $J$ and $K$-bands,
by inserting stars with known position and magnitude into the images, 
and then repeating the PSF photometric analysis as was used for the real stars (see Section \ref{optical}). 
The luminosity distribution of artificial stars is chosen in such a way that more stars were  inserted toward the fainter magnitude bins. 
The fraction of recovered artificial stars per bin of magnitude gives an estimate of the photometric completeness.
The photometry in the $V$-band is 100\% complete down to 21 mag and is reduced to 70\% at
23 mag. In  $J$-band,  the photometry is 100\% complete down to 17  mag  and  is reduced to $\sim$  90\% at 17-18 mag and $\sim$ 80\% at
18-19 mag. In $K$-band, 100\% completeness down to 16 mag, $\sim$ 90\% at 16-17 mag and $\sim$ 80\% at 17-18 mag.
The completeness estimated from  the histogram distribution of $J$-band detection (see \citealt{jose2016} for details) 
is  in  agreement with that of the artificial star method.  The completeness measurements of $J$ and $K$-bands agree with
the 90\% completeness limits mentioned in the UKIDSS GPS catalog details (\citealt{lucas2008}; $K$=18.0, $H$=18.75, $J$=19.5 mag). 
The histogram distribution of sources in 3.6 and 4.5 $\mu$m bands shows that 90\% completeness corresponds to 15 mag in both bands.
The completeness limits measured for various bands are listed in Table 1.

\section{The pre-main sequence population  and cluster properties}

\subsection{Extinction towards the cluster}
\label{extinction}

We  derive the  extinction towards the cluster  using the $H-K$ colors of the background stars,
following the method  outlined by  \citet{gutermuth2005} (see also \citealt{jose2016} for further
details for our implementation). Briefly, after eliminating the foreground contribution,  the line-of-sight
extinction to each point on a 10$^{\prime\prime}$  $\times$ 10$^{\prime\prime}$  grid is estimated by evaluating the mean and standard 
deviation of the $H-K$ colors of the 20 nearest stars to the grid point.   An iterative outlier $H-K$
color algorithm is used to reject the colors more than 3$\sigma$ from the mean until the mean converges. 
The visual extinction map is generated assuming an average $H-K$ color of 0.2 mag for background stars and using the following relation, 
$A_V$ = 15.93$[(H-K)_{obs}  - (H-K)_{intrinsic}]$, assuming the reddening law by \citet{cardelli1989}. 
We assume $R_V$ = 3.1, because  Paper 1 shows  that  normal interstellar reddening law is applicable within the cluster.

%###########################
\begin{figure}
%\centering
\includegraphics[scale = 0.5, trim = 20 0 0 0, clip]{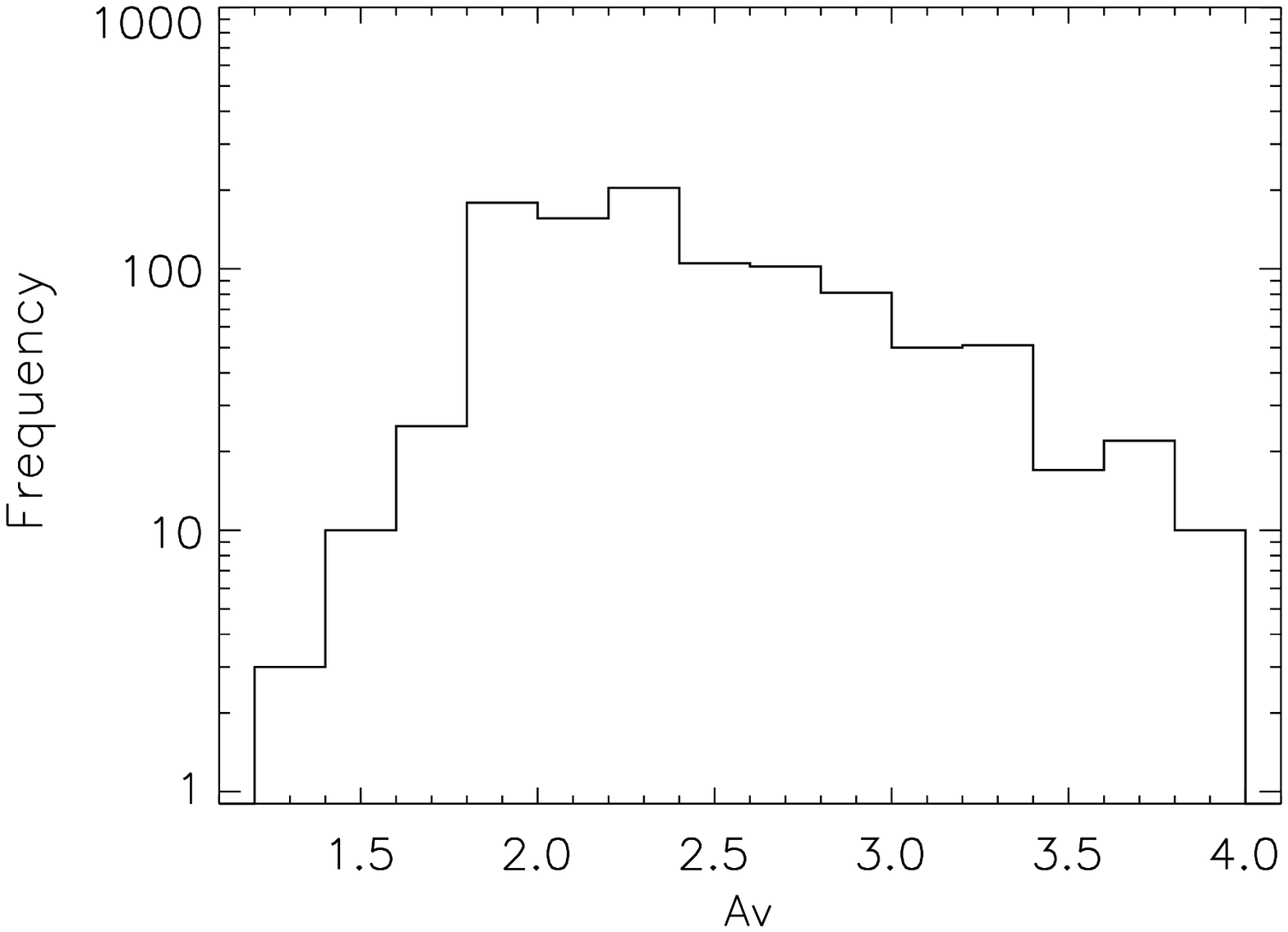}
\caption{Histogram distribution of extinction in $V$-band within 3$^\prime$ radius of the cluster. }
\label{ext}
\end{figure}
%%%%%%%%%%%%%%%%%%%%%%%%%%%%%%%%%%%%%%%%%%%%%%%%%%%%%%%%%%%%%%%%%%%%%%%%%%%%%%%%%%%%%%%%%%%%

The histogram  distribution of  extinction in  $V$-band  within the cluster is shown in Fig. \ref{ext}.
The median  extinction  of the  cluster is $\sim$ 2.0   mag  with a standard deviation of   $\sim$ 0.4 mag,  
is consistent with other extinction measurements towards the cluster (Paper I; \citealt{green2015,marco2016}).
The relatively low reddening variation within the cluster may be caused by the strong stellar winds of the  massive stars
around the cluster (Paper 1; \citealt{marco2016}). We use the extinction map
to deredden the point sources for YSO selection (Section \ref{yso}).

\subsection{Identification of young stellar objects}
\label{yso}
%###########################
\begin{figure*}[t]
\centering
\includegraphics[scale = 0.7, trim = 0 0 0 0, clip]{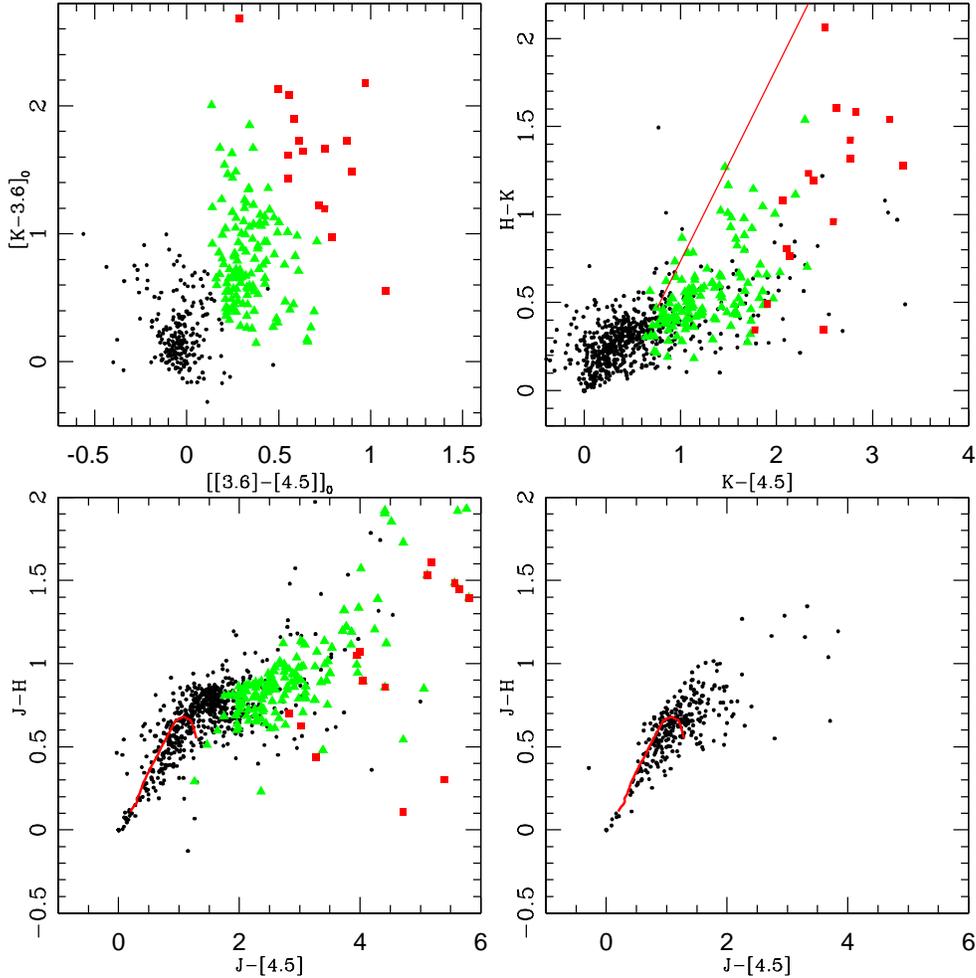}
\caption{NIR+IRAC color-color diagrams in various combinations for  all the sources within 3$^{\prime}$ radius of the cluster (top
panels and bottom left panel). Red and green are the candidate  Class I  and Class II sources. The bottom right panel
shows the distribution of sources in the control field region for an area equivalent to 3$^\prime$ radius. 
The red straight line represents the reddening vector drawn from the tip of the dwarf locus 
and the red curve in the bottom panels are the dwarf locus given by \citet{pecaut2013}.  }
\label{cc}
\end{figure*}
%%%%%%%%%%%%%%%%%%%%%%%%%%%%%%%%%%%%%%%%%%%%%%%%%%%%%%%%%%%%%%%%%%%%%%%%%%%%%%%%%%%%%%%%%%%%

In this section, we identify and classify the candidate Class I and Class II  young stellar objects
within Stock 8 based on their color excess in  the $J,H,K$, 3.6 and 4.5 $\mu$m bands.  We do not classify the diskless Class III sources
because they are indistinguishable from the field stars in their IR colors. However, in Section \ref{decontamination},
we estimate the probable statistics of the Class III population after removing the field star contaminants. 

In the first step, we use the color excess in $H$, $K$, 3.6 and 4.5 $\mu$m bands to identify the YSOs  following
the various color relations given by \citet{gutermuth2009}. We deredden the individual point sources based on their 
location on the extinction map  (see Section \ref{extinction}) and using the extinction law of 
\citet{flaherty2007}.  Within a 3$^\prime$ radius, 150 candidate YSOs (133 Class II and 17 Class I) are identified.   
The dereddened $[3.6-4.5]$ vs $K-[3.6]$  color-color diagram 
is shown in Fig. \ref{cc}, where the Class I and Class II sources are shown in red and green colors, respectively.

We next use a combination of $H$, $K$ and 4.5 $\mu$m bands to identify  additional YSOs. 
Fig. \ref{cc} (top right panel) shows the $K-[4.5]$ vs $H-K$ color-color distribution of all sources detected 
along with the 150 candidate YSOs already identified. The reddening vector from the tip of the dwarf locus 
is also shown. Following the methods of \citet{samal2014} and \citet{jose2016}, those sources with $H-K$ $ >$ 0.65
mag and with an excess $>$ 3$\sigma$  (where $\sigma$ is the average uncertainty in color) from the reddening vector  are considered 
as YSOs (see Fig. \ref{cc} top right panel). These sources also satisfy the color-color criteria to identify  the YSOs 
(i.e., $K - [4.5]$ $>$ 0.49 mag and $J - H$  $>$ 0.7 mag) given by \citet{zeidler2015}. We add 37 more YSOs to the list. 
Since most of these newly identified IR excess sources fall at the locations of the above {\it Spitzer} identified Class II YSOs, we consider them
as Class IIs, although a few could be Class I YSOs. Thus our catalog includes a total of 187 candidate YSOs within 3$^\prime$
radius of Stock 8, of which 147 are detected in $V$ and $I$ bands.  

The bottom left panel of  Fig. \ref{cc} shows the $J-[4.5]$ versus $J-H$ color-color diagram of the cluster region
along with the candidate YSOs. The bottom right panel shows that of a nearby control
field region, for an area equivalent to 3$^\prime$ radius, located  $\sim$ 15$^\prime$ south-east  
to the cluster, well outside the cluster boundary and  devoid of any $K$-band nebulosity.  A comparison of the
color-color distribution of the candidate YSOs  in  the bottom left panel of Fig. \ref{cc} with that of the
control field region shows that all the selected YSOs are likely IR excess sources. Since the intra-cluster 
reddening variation is negligible (see Section \ref{extinction}), the color excess of these YSOs is very unlikely to 
have any contribution from  extinction within the cluster. A comparison of the cluster and control field color-color
diagrams shows that $<$ 10\% candidate YSOs overlap with the location of field stars, which can be considered as the  
contamination from various sources (e.g., AGB stars, galaxies etc.).

\subsection{Field star decontamination and probable Class III statistics}
\label{decontamination}

%###########################
\begin{figure}
\centering
\includegraphics[scale = 0.55, trim = 0 0 0 0, clip]{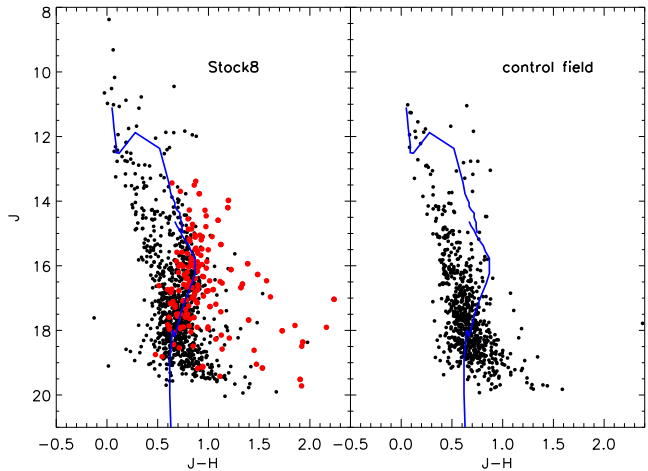}
\caption{$(J-H)/J$   color-magnitude  diagram  for  all the sources (black) within 3$^{\prime}$ radius of the cluster 
({\it left}) and for sources within the control field region of same area ({\it right}).
The red sources are the candidate Class I and Class II YSOs identified in Section \ref{yso}. The  pre-main sequence isochrone for
3 Myr (from \citealt{siess2000} and \citealt{baraffe2015}) are shown by the blue curves.  }
\label{cmd_jhj}
\end{figure}
%%%%%%%%%%%%%%%%%%%%%%%%%%%%%%%%%%%%%%%%%%%%%%%%%%%%%%%%%%%%%%%%%%%%%%%%%%%%%%%%%%%%%%%%%%%%

%###########################
\begin{figure*}[t]
\centering
\includegraphics[scale = 0.8, trim = 0 0 0 0, clip]{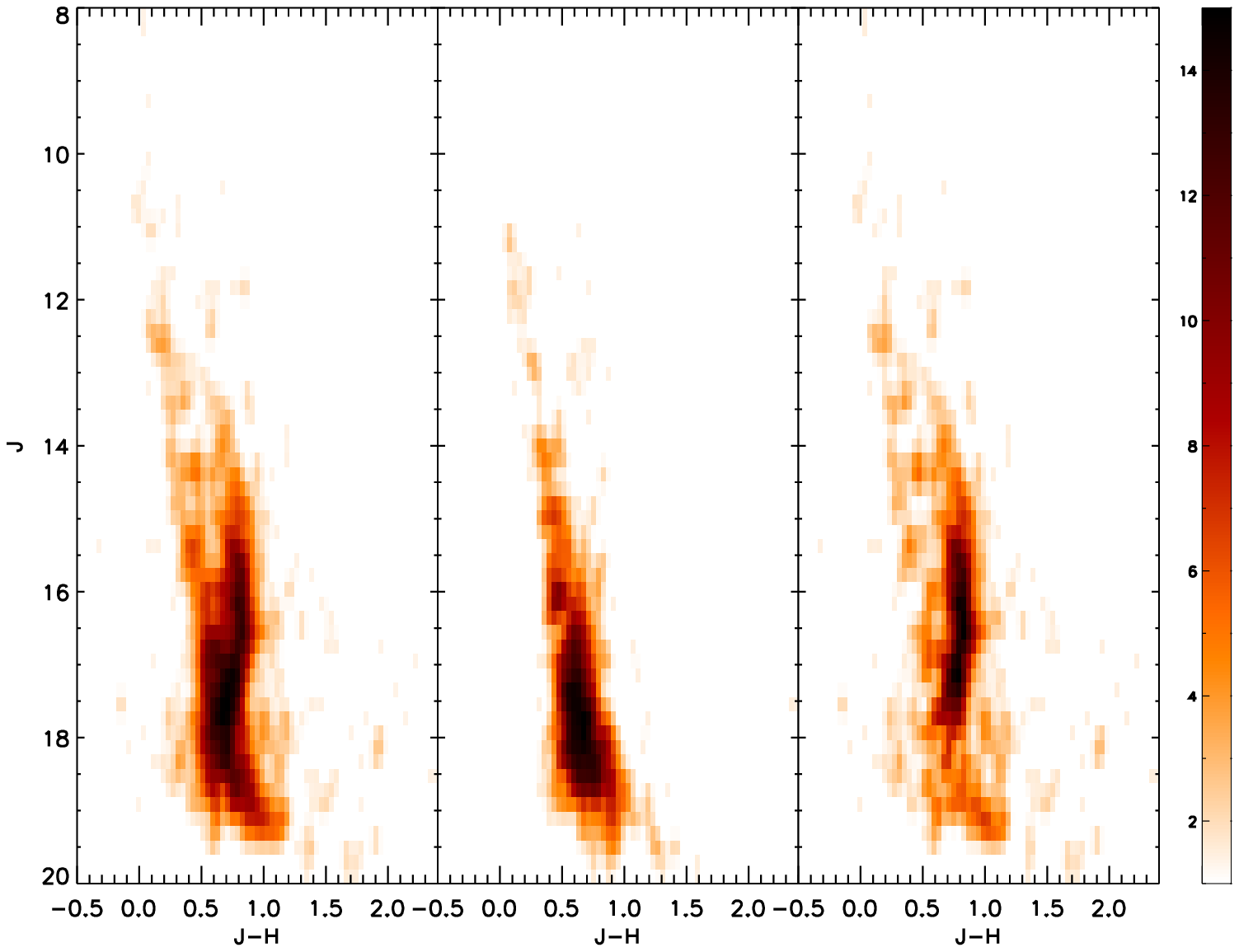}
\caption{ Hess diagrams of the $(J-H)/J$ stellar density in the cluster (left), for the  control field of similar area (middle)
and after the  field star decontamination (right). Two distinct stellar sequences apparent in the left  panel correspond to the
field star/MS and the pre-main sequence of the cluster members.  }
\label{nir_cmd}
\end{figure*}
%%%%%%%%%%%%%%%%%%%%%%%%%%%%%%%%%%%%%%%%%%%%%%%%%%%%%%%%%%%%%%%%%%%%%%%%%%%%%%%%%%%%%%%%%%%%

The left panel of Fig. \ref{cmd_jhj}  shows the $(J-H)/J$ color magnitude diagram (CMD) of the stars within 
3$^\prime$ radius of the cluster along with the  pre-main sequence (PMS) isochrones of 3 Myr from \citet{siess2000} ($>$ 1.4 $M{_\odot}$)
and \citet{baraffe2015} ($<$ 1.4 $M{_\odot}$). Candidate Class I/II YSOs  identified in Section \ref{yso} are also overplotted in the figure. 
The right panel of Fig. \ref{cmd_jhj}  shows the $(J-H)/J$ distribution of sources within the control field region. 
A comparison of two figures shows that the sequence of stars seen on the left side in both figures  is generally the field
star population of the region and the extra sequence to the right  is   the locus of candidate pre-main sequence cluster members.

In the left panel of Fig. \ref{cmd_jhj}, the pre-main sequence is well defined by the YSOs and lies well separated  from the MS/field star 
distribution of the cluster between $\sim$ 14-17 mag in $J$. The pre-main sequence merges with the  field star distribution at fainter levels.  
In order  to study the circumstellar disk fraction or luminosity/mass function,  
field stars should be removed  from the sample of stars in the cluster region.
 Without proper motion or spectroscopic data  for the cluster, statistical subtraction 
of the control field from the cluster region is the only way to
determine the member stars. Since the reddening towards Stock 8 and the control field is comparable, we assume
that the same background stars exist in the cluster region as that in the control field. We 
statistically remove the  expected background sources as follows.
We first divide the $J/(J-H)$ space in to several  grids of sizes 0.125 and 0.065 mag in the $J$ and $J-H$  axes both
for the cluster and field CMDs. For any star in the $J/(J-H)$  CMD of the control field region, the nearest star in the 
cluster  $J/(J-H)$ CMD  within the same grid was considered as a field star and removed.
By repeating this procedure for all sources in the CMD, we obtain the  background subtracted  sources in the target 
field and $\sim$ 700 sources are cleaned within 3$^\prime$ radius. We repeat the process using  two other
control field regions located to the north-east and south-west of the cluster  and the statistics of the
field decontaminated CMD remains same within Poisson uncertainty. This  statistical
cleaning leaves 556 sources within 3$^\prime$ radius, including the Class I/II/III population. 
In Section \ref{yso}, we identified  187  candidate Class I/II sources, so  there are approximately $\sim$ 369 (556-187)
 likely candidate Class III sources in the cluster. 

 In order to study  the spread  of the pre-main sequence and field star population, in Fig. \ref{nir_cmd} we show
 Hess\footnote{A Hess diagram shows the relative density distribution of stars at different colour-magnitude bins on the HR diagram} diagrams 
as a way to represent the stellar density distribution in $(J-H)/J$ axes. We bin 
the stars in the color and magnitude axis of the CMD and smooth the 
counts with a Gaussian kernel.    Fig. \ref{nir_cmd}  shows the $(J-H)/J$ Hess diagrams for the sources in
the  cluster region (left panel), the control field (middle panel), and the final result after the background elimination
(right panel), respectively. Two sequences for the field and pre-main sequence population are evident in the left panel 
and  majority of the field population have been removed in the field decontaminated CMD shown in the right panel. 

\section{ HR diagram analysis : Ages and age spread of young stellar objects}
\label{age}

%###########################
\begin{figure*}[t]
\centering
\includegraphics[scale = 0.7, trim = 0 0 0 0, clip]{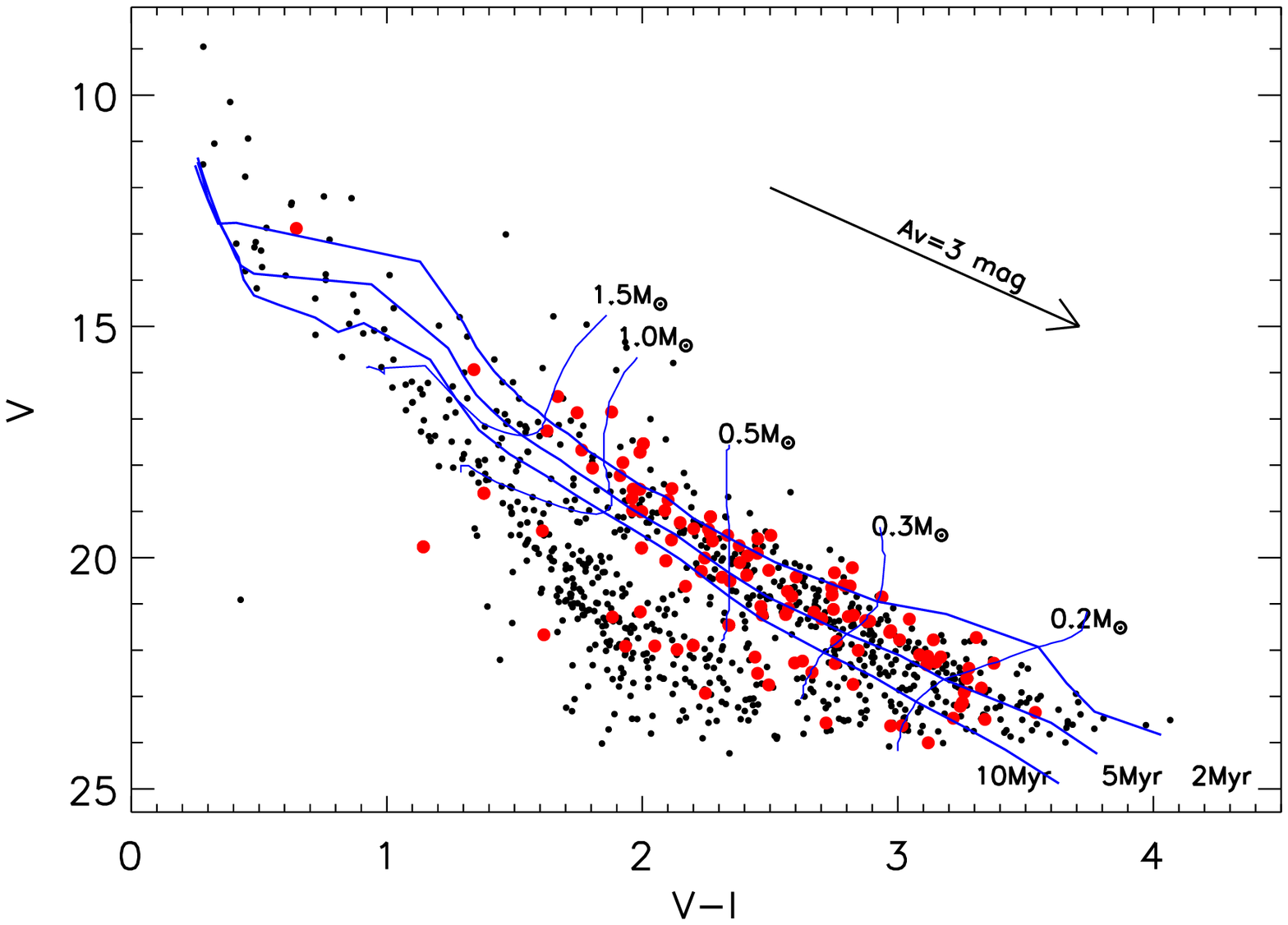}
\caption{$(V-I)/V$   color-magnitude  diagram  for  all the sources (black) within 3$^{\prime}$ radius of Stock 8. 
The red sources are the candidate Class I/II YSOs identified in Section \ref{yso}. The blue curves are the
pre-main sequence isochrones for 2, 5 and 10 Myr from \citet{siess2000}, and evolutionary tracks (thin solid curves) for various masses, which are
corrected for the  cluster distance and reddening. Reddening vector for $A_V$ = 3 mag is also shown.  }
\label{cmd}
\end{figure*}
%%%%%%%%%%%%%%%%%%%%%%%%%%%%%%%%%%%%%%%%%%%%%%%%%%%%%%%%%%%%%%%%%%%%%%%%%%%%%%%%%%%%%%%%%%%%

%%%%%%%%%%%%%%%%%%%%%%%%%%%%%%%%%%%%%%%%%%%%%%%%%%%%%%%%%%%%%%%%%%%%%%%%%%%%%%%%%%%%%%%%%%%%
\begin{figure}
\centering
\includegraphics[scale = 0.9, trim = 0 0 0 70, clip]{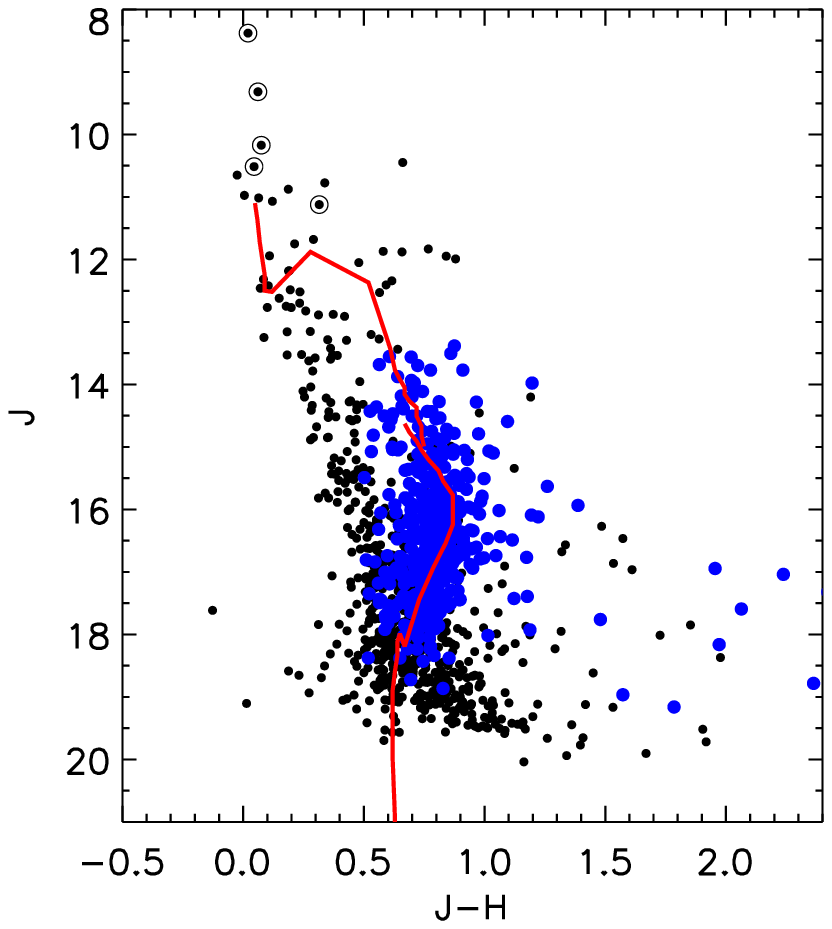}
\caption{$(J-H)/J$   color-magnitude  diagram  for  all the sources (black) within 3$^{\prime}$ radius of the cluster.
  The blue sources are those sources lying above 10 Myr isochrone in Fig. \ref{cmd} and with mass $<$ 1.5 $M_{\odot}$.
  The curves are the pre-main sequence isochrones for 3 Myr. The five encircled sources at the
upper main-sequence are the massive stars identified within the  radius by \citet{marco2016}.
}
\label{jhj_cmd}
\end{figure}
%%%%%%%%%%%%%%%%%%%%%%%%%%%%%%%%%%%%%%%%%%%%%%%%%%%%%%%%%%%%%%%%%%%%%%%%%%%%%%%%%%%%%%%%%%%%

The distribution of YSOs on the optical  CMD can be used  to estimate their approximate age. 
The relatively low and constant  reddening detected  towards Stock 8 (Section \ref{extinction}) is an advantage to examine the position of YSOs 
on the CMD  and to estimate their approximate  age and mass based on the evolutionary models.
Fig. \ref{cmd} shows the $(V-I)/V$ CMD for  sources within a 3$^\prime$ radius of Stock 8.
The candidate YSOs identified based on  IR excess  (see Section \ref{yso}) are also superposed in the figure.  
A discontinuity in the stellar color distribution in the CMD  separates the  population of  MS/field
stars   from one grouped at redder colors by 1--2 mag. Most candidate YSOs are located along   the redder  group of stars.
The CMD shown in  Fig. \ref{cmd} is $\sim$ 2  mag fainter than those studied in Paper 1.

Cluster ages are generally inferred by fitting the distribution of  young stellar population  with pre-main sequence evolutionary
model grids. The pre-main sequence isochrones of age 2, 5 and 10 Myr from  \citet{siess2000}   are overplotted
in Fig. \ref{cmd} after correcting for the  distance and minimum reddening (i.e., $A_V$ = 1.2 mag; Paper 1; \citealt{marco2016})
towards the cluster. Fig. \ref{cmd}  also shows the pre-main sequence evolutionary tracks  for various masses.
The  YSOs are  mostly distributed  in the mass range between 0.2 $M_{\odot}$ - 1.5 $M_{\odot}$.

In  Fig. \ref{cmd}, other than the candidate YSOs, many sources are located above the  pre-main sequence isochrone  of 10 Myr.
These sources are likely to be the candidate Class III sources that could not be selected in our YSO list based on IR excess. 
Fig. \ref{jhj_cmd} shows  the $(J-H)/J$ distribution of all those sources located above 10 Myr isochrone  of Fig. \ref{cmd} and $\le$ 1.5 $M_{\odot}$.
These sources  are mostly distributed along the pre-main sequence locus in Fig. \ref{jhj_cmd}, similar to the field-decontaminated
CMD shown in Fig. \ref{nir_cmd}.
The statistics of those sources located above 10 Myr isochrone  in Fig. \ref{cmd} and above the optical completeness limit (mass $>$ 0.3 $M_{\odot}$)
is similar to  that of the field-decontaminated CMD in Fig. \ref{nir_cmd}, within Poisson uncertainty.
This shows that the statistics of the candidate Class III sources estimated in Section \ref{decontamination} is genuine.

%###########################
\begin{figure}
\centering
\includegraphics[scale = 0.35, trim = 10 10 10 0, angle=-90, clip]{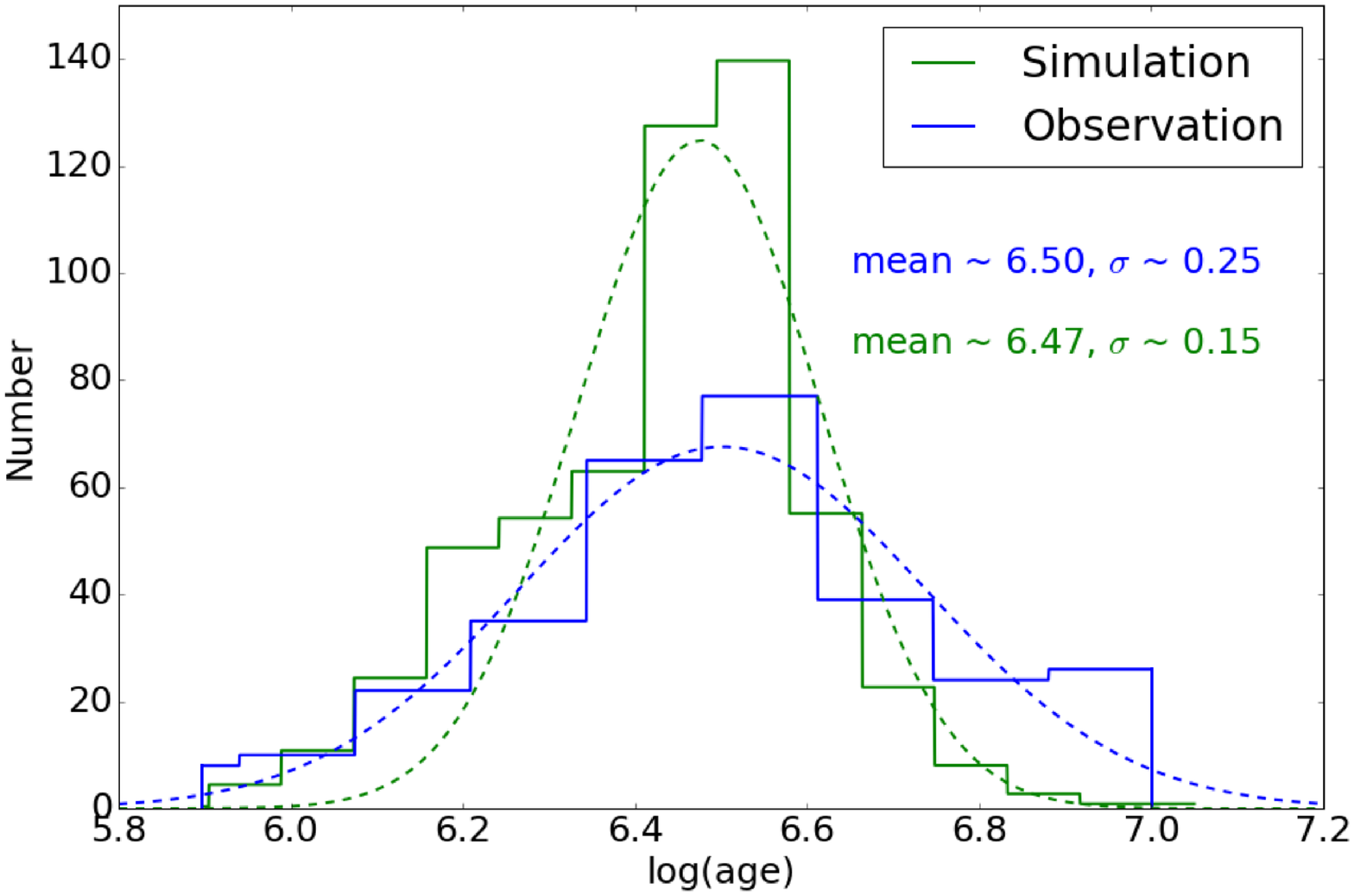}
\caption{Age distribution of sources  located above  10 Myr isochrone in the $(V-I)/V$ color-magnitude diagram (Fig. \ref{cmd}) and
  mass range of  1.5 -- 0.3 M$_\odot$ (blue). The apparent age spread of the  simulated color-magnitude diagram after propagating the age spread due to
  various factors  and scaled to the number of YSOs (see text for details) is shown in green. The mean and sigma of the distributions are also shown. 
}
\label{hist_age}
\end{figure}
%%%%%%%%%%%%%%%%%%%%%%%%%%%%%%%%%%%%%%%%%%%%%%%%%%%%%%%%%%%%%%%%%%%%%%%%%%%%%%%%%%%%%%%%%%%%

We estimate the age of  the individual cluster members  by comparing their locations on the CMD with 
the grid of pre-main sequence isochrones of ages  0.1 Myr to 10 Myr \citep{siess2000}. 
Fig. \ref{hist_age}  shows the histogram distribution of ages of all the sources  in
the range 1.5 -- 0.3 M$_\odot$  and located above the 10 Myr isochrone in the  $(V-I)/V$ color-magnitude diagram (Fig. \ref{cmd}). 
The Gaussian fit to the distribution gives a mean log(age) $\sim$ 6.5 ($\sim$ 3 Myr) and $\sigma$ of $\sim$ 0.25 dex.
The 1$\sigma$ age spread for Class I+II sources is $\sim$ 0.2 dex and that for Class III sources is $\sim$ 0.3 dex. 

\subsection{Age of the cluster from disk fraction}

The fraction of circumstellar disk sources in a star forming region can be used as a proxy for its age estimation 
(e.g. \citealt{haisch2001,hernandez2008,fang2013}). 
In order to estimate the disk fraction of Stock 8, we use the field-decontaminated source list (see Section \ref{decontamination})
as the total member stars (i.e., disked and  disk-less) within the cluster. Our YSO survey is limited by the sensitivity
of the IRAC observations i.e., 15 mag in 3.6 and 4.5 $\mu$m bands (see Section \ref{completeness}). 
 The corresponding photospheric brightness in $J$-band is 16.5 mag. Hence we consider only those sources brighter than 16.5 mag in $J$,
for estimating  the disk fraction, which corresponds to $\sim$ 0.4 $M{_\odot}$ at 3 Myr \citep{baraffe2015}.
Sources within the 1.5 - 0.4 $M{_\odot}$  mass range  have  a disk fraction of  $\sim$ 35\% in  Stock 8.
The average disk fraction versus age trend reported in the literature (e.g. \citealt{haisch2001,hernandez2008,fang2012,fang2013})
in JHKL bands suggests, $\sim$ 50\% of the stars in a given region should have lost their disks around 2-3 Myr.
The age estimation based on disk fraction within Stock 8 is consistent  with the photometrically derived age
from the optical CMD analysis.

\subsection{Probable causes of the apparent age spread}

The age estimation of  any star forming region is usually plagued with uncertainties (e.g. \citealt{hartmann2001,soderblom2014}).
In addition to the  uncertainties in the input physics affecting the evolutionary models,  various observational factors
 could cause an incorrect estimation  of the age (or mass) of individual objects and  inference of  accurate star formation histories of 
molecular clouds \citep{hartmann2001,slesnick2008}. The typical sources of  observational uncertainties include photometric errors,
unresolved binaries, differential reddening, the modulation of spots, the obscuration by disk material  
 and accretion  (Stauffer et al. 2016). All of these effects produce  broadening in the CMD, which could be 
misinterpreted as true age spread. Below we explain the apparent age spread in the cluster  by generating a synthetic CMD and by broaden  it
with  the individual factors, which can cause the major broadening in the CMD (i.e., Photometric uncertainty, differential reddening, binarity
and variability).  We generate a synthetic CMD using 10000 artificial
pre-main sequence  stars in the mass range of 0.3 to 1.5  $M_{\odot}$ with the  models of \citet{siess2000}.
The underlying IMF is assumed to be the Salpeter IMF \citep{salpeter1955}. The age of the artificial stars is set to 3 Myr 
and an instantaneous star formation history is assumed, i.e. no spread in the age. The age spread due to individual factors is 
applied to the simulated CMD  using Monte-Carlo simulations to estimate the effective spread in age. 

{\it Photometric uncertainty}: 
The median value of photometric uncertainty  within the completeness limit is $\sim$ 0.04 and 0.05 mag in $V$ and $V-I$, respectively.
This uncertainty is mainly caused by the PSF profile fitting and is random. We broaden the synthetic CMD with 
 the above median values of  photometric uncertainty and the corresponding age spread  in log(age) is  $\sim$ 0.09 dex. 
 
{\it Differential reddening}:
In Section  \ref{extinction}, we derive the mean reddening of the cluster as $\sim$ 2 mag with a standard deviation
of $\sim$ 0.4 mag in $V$-band and $\sim$ 0.16 mag in $V-I$.  After broadening the CMD for a differential  extinction of $\sim$ 0.4 mag
in $V$ and 0.16 mag in $V-I$, we obtain the corresponding age spread in  log(age) as  $\sim$ 0.07 dex.

{\it Binarity}:
We generate   binary stars based on the empirically derived relations for  mass-dependent multiplicity fraction   by  \citet{lu2013}
and assuming a flat mass-ratio distribution. The standard deviation of the age spread  in the synthetic CMD due to binarity is $\sim$ 0.13 dex.

{\it Optical variability}: Pre-main sequence stars can cause the spread in the observed CMD due to its variable nature.
The variability in  $V$-band for classical T Tauri stars (CTTSs) can be as  high as  3 mag, while for weak-line T Tauri stars (WTTSs)
the variability is $<$  0.6 mag  (\citealt{herbst1994,hennekemper2008}). The main source of variability in WTTSs is the  rotational
modulation of cold spots. Majority of the pre-main sequence  sources in Stock 8 are WTTSs.  We broaden the synthetic CMD based on
the rms distribution of $r$-band variability observed in several hundreds of  WTTSs  by  \citet{venuti2015} and we obtain
standard deviation of the age spread in the synthetic CMD  $\sim$ 0.02 dex.

 The synthetic CMD  scaled to the number of observed YSOs and broadened by the above factors is shown in Fig. \ref{hist_age}. 
The synthetic CMD  has  an apparent  spread of $\sim$  0.15  dex in log(age), whereas the  observed age spread is $\sim$ 0.25 dex (Fig. \ref{hist_age}).
After quadratically subtracting the apparent age spread  from the observed spread,  we   obtain  the  true age spread
in the cluster as   $\sim$ 0.2 dex  in log(age).   The mean age and age spread of Stock 8 is 6.5 $\pm$ 0.2 dex or
  3.16 (+1.85 - 1.16) Myr.  The contribution from the true age spread due to star formation history
  of the cluster  is comparable  to other star forming regions such as W3-main (2-3 Myr; \citealt{bik2012})
  and NGC2264 (3-4 Myr;  \citealt{lim2016}). The  luminosity spread  corresponding to the 1$\sigma$ observed age spread at $V-I$= 2.5 mag is 0.24 dex. 

\section{Luminosity function and initial mass function of Stock 8}

In this section we use  the cluster's luminosity function  to derive  the initial mass function, which are the
two  fundamental parameters of a cluster to understand  its formation. 
Based on the photometry  completeness limits and age of the cluster,  the mass detection limit is  derived using
pre-main sequence  isochrones. Assuming the median age of the cluster  $\sim$ 3 Myr, median extinction   $A_V$ $\sim$  2.0 mag
(see Section \ref{extinction}), at a distance of 2.1 kpc, the photometric completeness limit of  J=19 mag for the cluster
(Section \ref{completeness}) corresponds to a mass limit of 0.05 $M_{\odot}$ (\citealt{baraffe2015}).  
Considering the uncertainty in  age estimation (Section \ref{age}), for the upper age limit of $\sim$ 5 Myr, the photometric
completeness limit  would raise the mass detection limit to 0.08 $M_{\odot}$ in $J$-band. 

The luminosity function is frequently used in studies of young clusters as a diagnostic tool of the 
mass function and the star formation history of their stellar populations (\citealt{lada2003}, and references therein). 
Pioneering work on the interpretation of the $K$-band luminosity function (KLF)  was presented by \citet{zinnecker1993}. Later,
models of the $K$-band luminosity function by \citet{muench2000} showed its dependence on factors such as the 
cluster age, the spread of star formation over time, and the choice of theoretical pre-main sequence  evolutionary tracks.

%###########################
\begin{figure}
\centering
\includegraphics[scale = 0.6, trim = 0 0 0 0, clip]{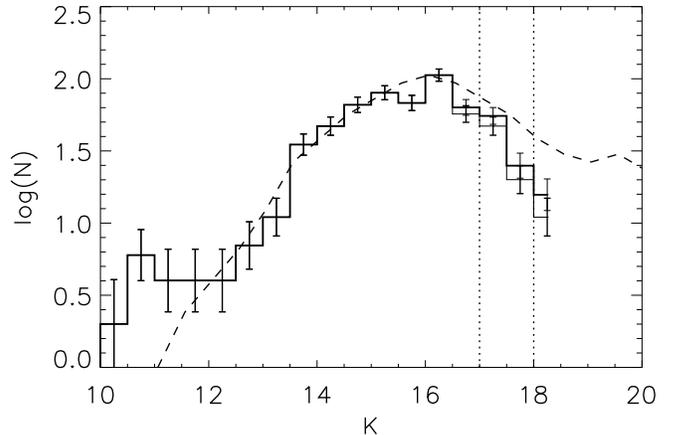}
\caption{$K$-band luminosity function of stars within the cluster radius (thin histogram) and after completeness correction (thick histogram).
The dashed curve represents the Trapezium KLF from \citet{muench2002} shifted for the distance of Stock 8 and  
scaled to match the peak of the KLF of Stock 8. The vertical dotted lines represent the 90\% and 80\% photometric completeness limits.  }

\label{klf}
\end{figure}
%%%%%%%%%%%%%%%%%%%%%%%%%%%%%%%%%%%%%%%%%%%%%%%%%%%%%%%%%%%%%%%%%%%%%%%%%%%%%%%%%%%%%%%%%%%

To derive the luminosity function, we use the $K$-band photometry of the field star decontaminated (Section \ref{decontamination})  data. 
 From the right panel of Fig. \ref{nir_cmd}, we derive a mean locus of the pre-main sequence stellar distribution in the $J-H/J$ plane, which
traces the 3 Myr isochrone of \citet{siess2000}  (also see Fig. \ref{cmd_jhj}). To derive the  luminosity function,
 we  exclude a few sources which fall beyond  3$\sigma$ from the mean locus of $J-H/J$ distribution. These sources are most 
likely field stars, that were not removed  due to the statistical uncertainty in our field 
star decontamination process. 

%###########################
\begin{figure}
\centering
\includegraphics[scale = 0.6, trim = 0 0 0 0, clip]{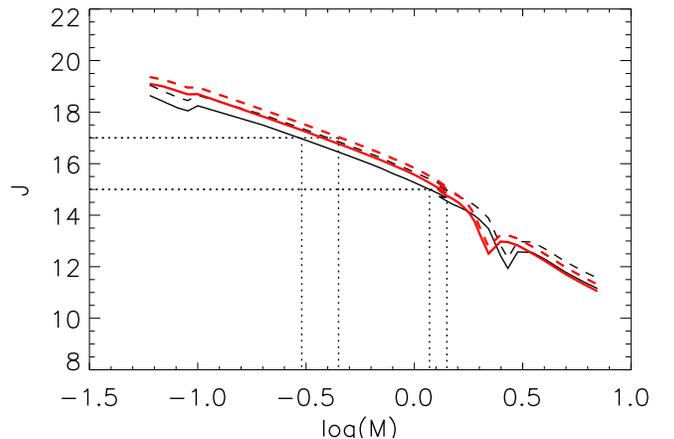}
\caption{  Mass-magnitude relation  used for the IMF estimation from \citet{baraffe2015} (for $<$ 1.4 M$_\odot$) 
and \citet{siess2000}  (for $>$ 1.4 M$_\odot$) models. The black curves are for 3 Myr and red are for 5 Myr age. The dashed curves 
are obtained  by incorporating uncertainty in extinction. The dotted lines show the expected error
in mass for a given luminosity.    }
\label{ml}
\end{figure}
%%%%%%%%%%%%%%%%%%%%%%%%%%%%%%%%%%%%%%%%%%%%%%%%%%%%%%%%%%%%%%%%%%%%%%%%%%%%%%%%%%%%%%%%%%%

Fig. \ref{klf} shows the  $K$-band luminosity function  of the cluster. The shape of the luminosity function appears
remarkably similar to the  well studied star forming regions such as Trapezium \citep{muench2002}, NGC2024 \citep{spezzi2015},
L1641 \citep{fang2012}, RCW41 \citep{neichel2015}, NGC6611 \citep{oliveira2009}. For comparison, the $K$-band luminosity function
of Trapezium cluster from \citet{muench2002}  is over-plotted in Fig. \ref{klf}, which shows a good match with Stock 8 up to
$K$ $<$ 17.5 mag, beyond which it deviates. It may be due to improper field subtraction or completeness correction at faint level.
In Fig. \ref{klf}, the luminosity function  peaks between $K$ $\sim$ 14.5--16.5 mag and below 16.5 mag the luminosity function
steadily declines to the hydrogen burning limit. 

%###########################
\begin{figure*}[t]
\centering
\includegraphics[scale = 0.4, trim = 0 0 20 0, clip]{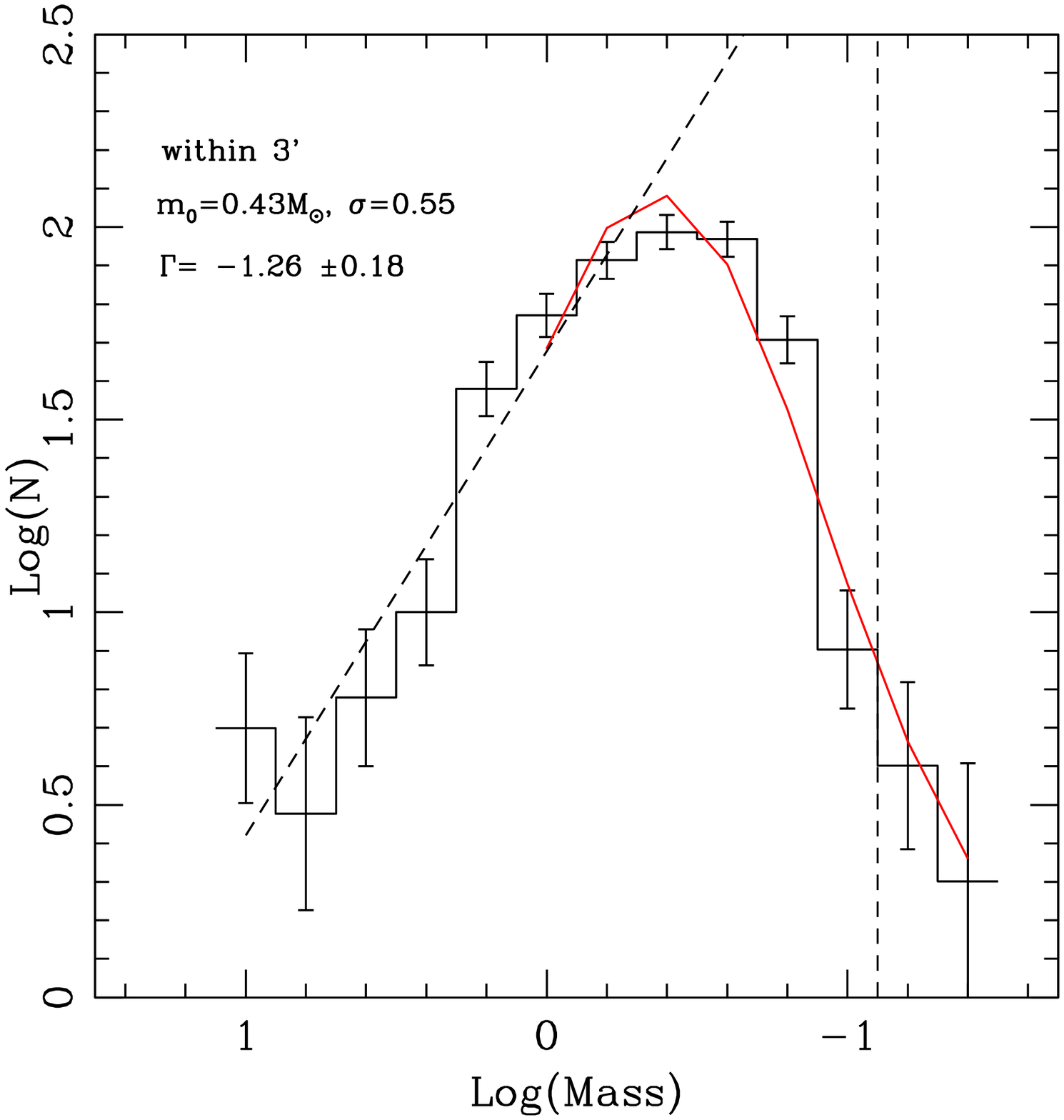}
\includegraphics[scale = 0.4, trim = 0 0 20 0, clip]{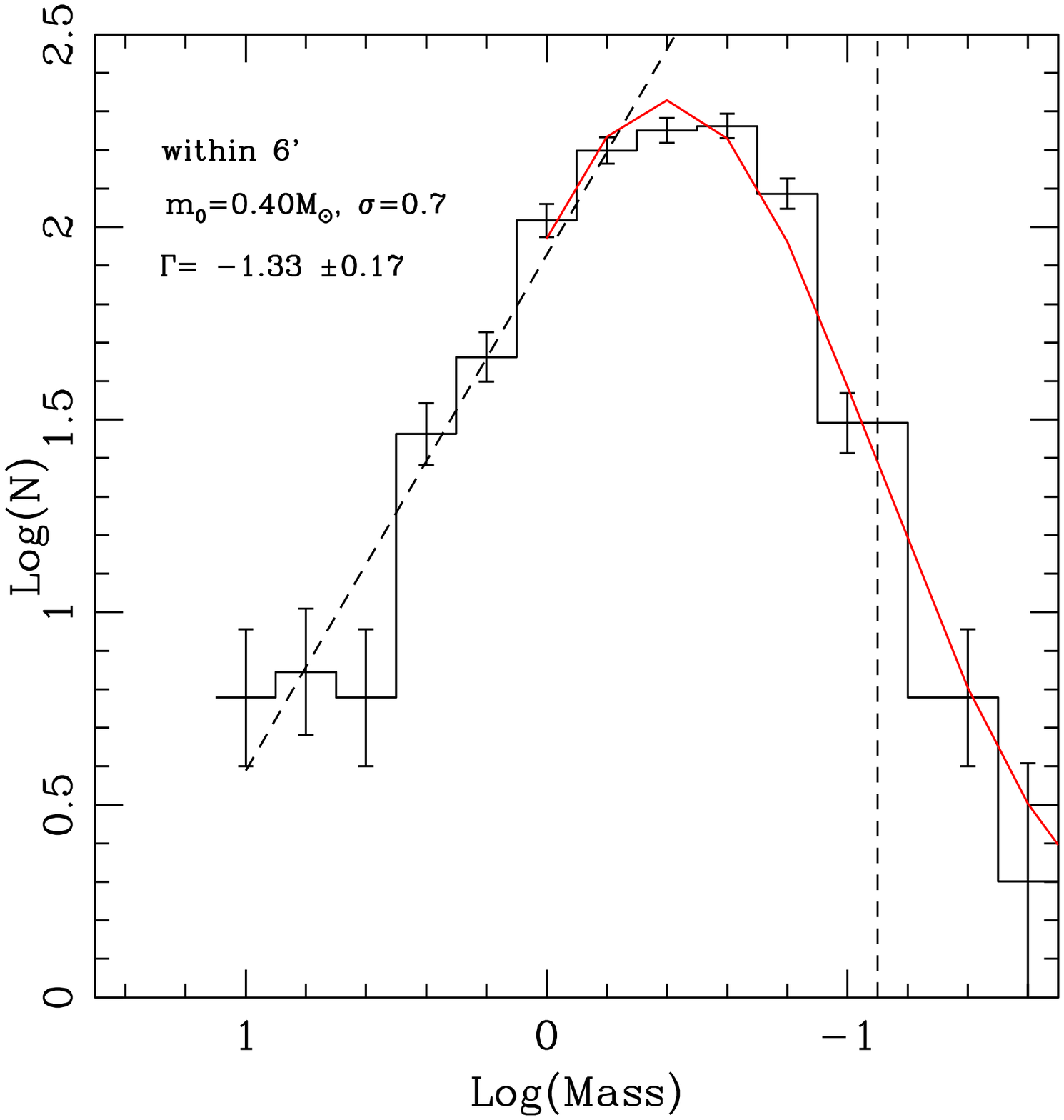}
\caption{The IMF derived from the $J$-band luminosity function after statistical subtraction of the control field sources,
 within 3$^\prime$ ({\it left}) and 6$^\prime$ ({\it right}) radius of the cluster. The \citet{chabrier2003} log-normal like distribution
 is shown by the red solid curve for $M<$ 1.0 $M_{\odot}$ and the slanted straight line represents the linear fit to the mass distribution for $M>$ 0.5 $M_{\odot}$.
 Vertical dashed line represents the 80\% completeness limit.   }
\label{mf}
\end{figure*}
%%%%%%%%%%%%%%%%%%%%%%%%%%%%%%%%%%%%%%%%%%%%%%%%%%%%%%%%%%%%%%%%%%%%%%%%%%%%%%%%%%%%%%%%%%%

The IMF is  useful  to understand  the formation and evolution of stellar systems and the contribution
of stellar mass to the host galaxy.  In order to estimate the mass of the individual stars in the field star decontaminated CMD
(see Fig. \ref{nir_cmd}), the mass-magnitude relation from stellar evolutionary models of 3 Myr is used.
For M $<$ 1.4 M$_\odot$, the 3 Myr PMS isochrone of \citet{baraffe2015}  and for  M $>$ 1.4 M$_\odot$,
\citet{siess2000} models are used for mass-magnitude conversion (see Fig. \ref{ml}).
For  younger ages ($<$ 10 Myr), the pre-main sequence isochrones are almost vertical.  
The color change (in $J-H$) of the YSOs  due to age variations within the  age range of $\sim$ 3-5 Myr
is minor since the effective temperature varies slowly. Since the reddening variation within the
cluster is minimal, the above assumptions  are reasonably valid to estimate the mass-magnitude relation.  
Fig.  \ref{ml} also shows the error in the mass estimation from the
mass-magnitude relation by incorporating the  uncertainty in the cluster age  ($\sim$ 2 Myr) as well as the uncertainty in the $A_V$
values ($\sim$ 0.4 mag). The typical uncertainty in log m is $<$ 0.15 dex.

IMF is calculated using the mass-magnitude relation and by counting  the number of stars in a 
logarithmic mass interval. The size of the IMF bin  set to be   log m = 0.2, which is comparable to the average uncertainty of
mass estimation propagated from the assumed mass-magnitude relation (see Fig. \ref{ml}). Fig. \ref{mf} (left panel)  represents
the mass distribution  of Stock 8 within 3$^\prime$ radius, which in general  follows the shape of the Galactic field MF. We find
that the IMF in Stock 8 rises Salpeter-like at high masses, consistent with many previous IMF estimations on young clusters
(eg. NGC2264; \citealt{sung2010}).  The most massive bin has 5 sources,  in agreement with the 5 early class stars identified within 3$^\prime$ radius
by \citet{marco2016}. The IMF flattens out between 0.2 - 0.7 $M_{\odot}$ and then drops down  to the brown dwarf regime. For $M$ $>$ 0.5 $M_{\odot}$,
 Salpeter  mass function is fit ($dN/d log(m)$ = $M^{\Gamma}$; \citealt{salpeter1955}), and is shown in Fig. \ref{mf}.
The value of the slope, $\Gamma$,  obtained ($\Gamma$ = -1.26 $\pm$ 0.18) agrees well with the canonical value (-1.35).  We also fit the mass
function with a log-normal distribution, $dN \over dlogm$ $ \propto$  exp(--$(m-m_0)^2 \over 2\sigma^2$), where, log m$_0$ is
the characteristic stellar mass 
(the  mass at peak of the distribution) and $\sigma$ is the spread in the  log-normal distribution.  The Galactic fieldmass function
has a characteristic mass $m_0$ = 0.25 $M_\odot$ and $\sigma$ = 0.55 $M_\odot$ \citep{chabrier2003}.
The derived log-normal functional fit to the low-mass stars (1.0 M$_\odot$ $>$ M $>$ 0.08  M$_\odot$) of Stock 8
(see Fig. \ref{mf} left panel) within 3$^\prime$ radius yields a peak mass of  $m_0$ = 0.43 $M_\odot$.

In order to assess the effect of binning, the above analysis is repeated using different combinations of mass limits and
shifting the mass bins in $\Delta$log m  by  0.05 and 0.1 as well as for different bin sizes.  The experiment is repeated for
field-decontaminated data using two different control fields.  From these experiments,  the difference obtained in the value of
$\Gamma$ is $\sim$ 0.25 dex. The characteristic stellar mass  varies within 0.2 $M_{\odot}$,  comparable
to  the uncertainty limits. The choice of various evolutionary models can affect the mass-luminosity relations and thus the form of IMF.
Use  PARSEC models (\citealt{bressan2012,chen2014})  instead of \citet{siess2000} models, yields a  similar slope and
characteristic mass.  In conclusion, for $M$ $>$ 0.5 $M{_\odot}$ within Stock 8, the mass function follows a power law with slope similar to the
Salpeter slope and the characteristic stellar  mass appears to be robustly constrained around 0.4 M$_\odot$.

Our optical and IRAC photometric survey is limited to 3$^\prime$ radius around Stock 8, however, the cluster spans  a radius of
6$^\prime$ (Paper 1).  We also calculate the IMF using NIR data for the entire cluster radius. After  field star decontamination, we obtain
952 probable cluster members. The mass function within 6$^\prime$ radius  is shown in the right panel of Fig. \ref{mf}. The slope of the
power law above 0.5 $M{_\odot}$ as well as the characteristic stellar mass remains same as that of  3$^\prime$ radius.

\section{Discussion}

The low-mass regime of the IMF has been the subject of numerous observational and theoretical studies over the past decade (see
\citealt{offner2014}). The higher mass stars mostly follow the  Salpeter  mass function \citep{salpeter1955}. At
lower masses, the IMF is less well constrained, but appears to flatten below  1 M$_{\odot}$ and exhibits a
turnover between  0.1 - 0.7 M$_{\odot}$, with fewer  stars of the lowest masses (\citealt{kroupa2002,chabrier2003}).
While the  higher mass domain is thought to be mostly formed through  fragmentation and/or accretion
onto the protostellar core (e.g., \citealt{padoan2002,bonnell2006}), in the low-mass and substellar regime additional physics is likely
to play an important role. The density, velocity fields, chemical composition,  tidal forces in the natal molecular clouds
and photo erosion in the radiation field of bright stars in the vicinity  can lead 
to different star formation processes and consequently some variation  in the IMF (\citealt{bate2005,bate2009,padoan2002,whitworth2004}). 
The characteristic mass is observed to be essentially constant for most star forming regions (see review by \citealt
{elmegreen2008}). However, numerical simulations relate the characteristic mass to the  thermal Jeans mass, and hence variation with respect to
environment is expected \citep{bate2005}.
 
In general, the IMF of Stock 8 is consistent with that of  other well studied young clusters such as IC348, the Trapezium, NGC2264, NGC 6611 and W3-main.
(\citealt{muench2003,luhman2000,sung2010,oliveira2009,ojha2009}). Within uncertainties, the peak mass of IMF for Stock 8 ($\sim$ 0.4 M$_{\odot}$)
is comparable to that of majority of young clusters such as IC348, NGC 6611 and RCW 41   (\citealt{luhman2003,oliveira2009,neichel2015}).

The total mass of the cluster is 362 M$_{\odot}$ and 583 M$_{\odot}$ within 3$^\prime$ and 6$^\prime$, respectively, after
integrating the IMF within the mass range 10 -- 0.08 M$_{\odot}$.  Even though we do not include brown dwarfs,
their contribution to the total mass of the cluster should be negligible. Stock 8 falls  in the  category of moderately massive cluster
(10$^2$ $M_\odot$ $<$ $M_{cl}$ $>$ 10$^3$ $M_\odot$) in the classification scheme of \citet{weidner2010}. Recently, \citet{pfalzner2016} found 
a correlation between the cluster mass, size and stellar mass  using  the clusters within the solar neighbourhood
($M_c$ =$CR{_c}^{\gamma}$, with $\gamma$ = 1.7 $\pm$ 0.2), which gives constraints for the theory of
clustered star formation. The mass estimate within the effective radius of Stock 8 follows the above relation and is
comparable to the nearby well known young clusters (\citealt{kuhn2014,pfalzner2016}).

\subsection{Large scale structure around Stock 8}

 Stock 8 is surrounded by  33 early type stars ($ > $ B3 V) within a 24 pc of its radius  \citep{marco2016}. 
\citet{kang2012} studied high-velocity  \hi gas at  {\it l}  $\sim$ 173$^\circ$  with the Arecibo 305-m telescope
(beam $\sim$ 3.4$^\prime$). The high velocity  \hi emission features are confined inside the radio continuum
filaments of a giant shell associated with the  \hii complex G173+1.5, (which is composed of five Sharpless \hii regions,
S231-235 at its peripheries). The giant shell is spatially correlated with Stock 8 and associated  \hii region  Sh2-234,
which are located at the southern boundary of the shell.  \citet{kang2012} estimated the expansion velocity of the shell as $\sim$ 55 km s$^{-1}$ and 
kinetic energy  of the shell as $\sim$ 2.5 $\times$ 10$^{50}$  ergs  for the assumed  distance of 1.8 kpc. The hard
X-ray emitting hot gas inside the shell has thermal energy of $\sim$ 3$\times$ 10$^{50}$ ergs. \citet{kang2012} argue  that
the large kinetic energy, high expansion velocity and hard  X-ray emitting hot gas  implies a supernova origin of the shell.  
The scenario  proposed is that  the  supernova expansion  could have triggered the formation of the OB stars currently exciting the \hii regions.  
Using trigonometric parallaxes for water masers in massive star-forming regions, \citet{choi2014} identified a cluster close to Stock 8 
(IRAS 05168+3634) with a parallax distance of 1.9 kpc and a radial velocity of $v_{LSR}$ = -15.5 $\pm$  1.9 km s$^{-1}$ \citep{sakai2012}.  
These values are consistent  with the $v_{LSR}$ measured for the S231-235 \hii complex  \citep{kang2012}. 
The \hii complex S231-235 including Stock 8 shares the same radial velocity and are   likely located in the  Perseus Arm \citep{choi2014}.

\begin{figure*}[t]
\centering
\includegraphics[scale = 0.68, trim = 50 30 120 10, angle=-90, clip]{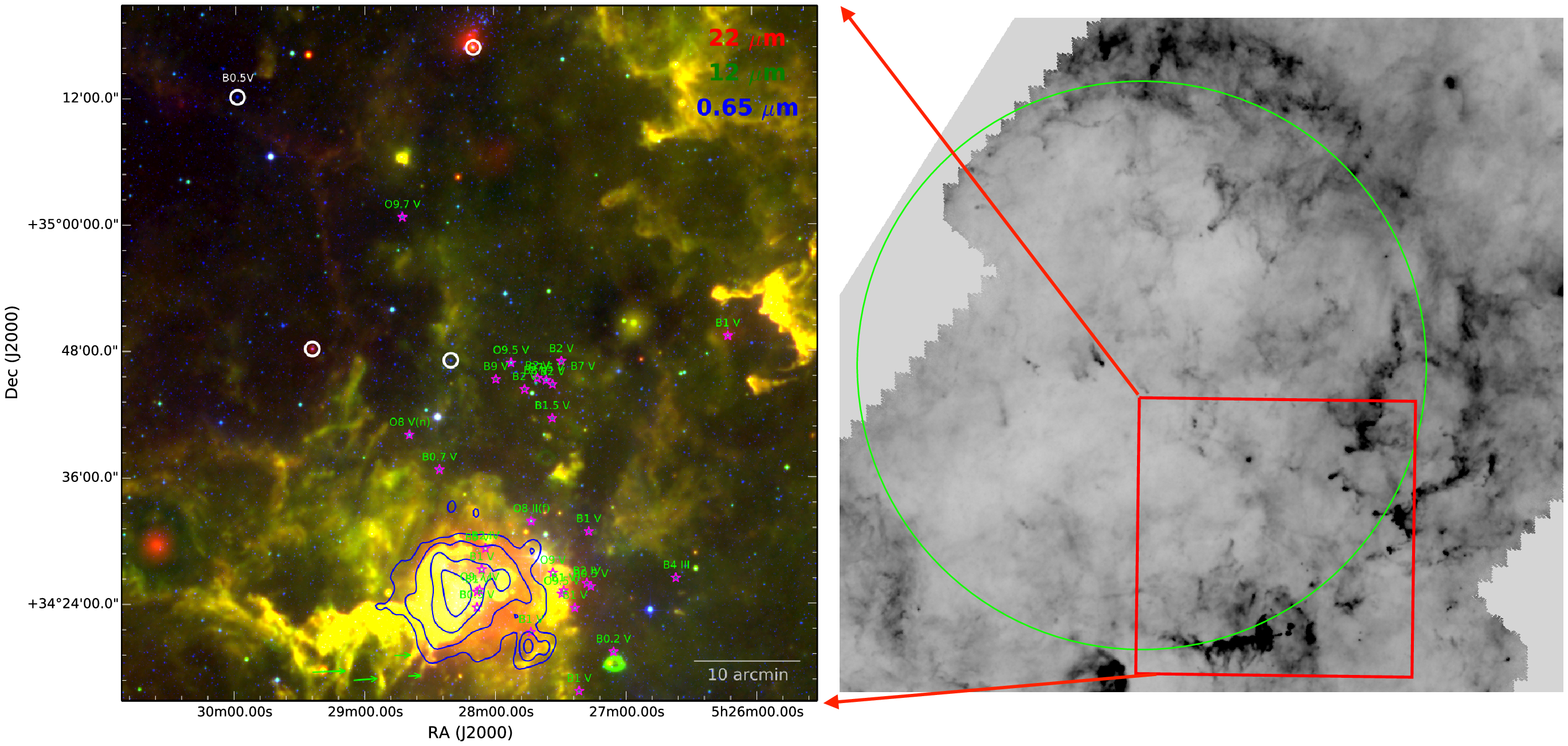}
\caption{  Color composite image made from DSS2 0.65 $\mu$m (blue), WISE 12 $\mu$m (green) and WISE 22 $\mu$m (red)
  images for Stock 8 and its surrounding  area ({\it left}) and {\it Herschel}-SPIRE image showing  the large scale
structure around Stock 8 ({\it right}).  The purple star symbols represent the locations of the
massive stars  from \citet{marco2016} and white circles represent those obtained from SIMBAD catalog.   Blue contours represent the
extended radio emission in 1420 MHz  obtained from CGPS \citep{taylor2003}.} 
\label{spatial}
\end{figure*}

 Fig. \ref{spatial} (right) shows the large scale structure around Stock 8, a large, 
irregular cavity at 350 $\mu$m (rough boundary is marked with a green circle) with little dust inside.
Stock 8 is located at its southern periphery. Fig. \ref{spatial} (left) 
shows the environment in the immediate vicinity of  Stock 8 at optical (DSS2-R; blue), 12 $\mu$m (WISE W2-band; green) 
and 22 $\mu$m (WISE W4-band; red).   The early type stars  identified by Marco \& Negueruela (2016)
as well as from SIMBAD are marked in Fig. \ref{spatial}. In Fig. \ref{spatial}, PAH emissions, which are excited by UV photons from
the massive star(s),  are present at the interface between the molecular and neutral gas. PAHs  are absent in the
interior of the shell due to destruction of these  molecules by intense UV radiation \citep{pomares2009}. 
The early type massive stars located around  Stock 8 should be responsible for this  12 $\mu$m cavity.
The irregular nature of the large cavity  could be caused by the combination 
of an inhomogeneous medium, presence of multiple OB stars, proper motion of the stars and projection effect along the line of sight. 

In Fig. \ref{spatial},  many  `pillar-like' or `elephant trunk' structures are projected inwards from the bubble walls towards
the OB-stars. Their heads point towards the bubble, and  filamentary tails lead away from the head.  High-resolution
observations and simulations have shown that these structures result from the  effect of radiation feedback on turbulent molecular cloud
(\citealt{gritschneder2010,jose2013,mcleod2015,mcleod2016,schneider2016,sharma2016}). Gravitational collapse occurs at the tip of
the pillars, leading to the formation of cores and low mass stars (\citealt{gritschneder2009,gritschneder2010,deharveng2012,chauhan2011}).  

%###########################
%%%%%%%%%%%%%%%%%%%%%%%%%%%%%%%%%%%%%%%%%%%%%%%%%%%%%%%%%%%%%%%%%%%%%%%%%%%%%%%%%%%%%%%%%%%

Stock 8 is located at the southern boundary of the cavity. Expanding super-bubbles may provide an efficient mechanism to compress
the low density ISM or pre-existing filaments into dense shells that become gravitationally unstable and form stars. The evidence
that  Stock 8 and its associated nebulous stream (see Paper 1 for a discussion) is being compressed can be seen from the sharp decline
of 12 $\mu$m emission at the inner edge of the bubble and from the presence of elongated striations   (marked using  green arrows in
Fig. \ref{spatial})  with their head pointing towards the center of the bubble. Newly formed OB stars (up to a few Myr old), generally
associated with strong radio and 24 $\mu$m emission, as found in several Galactic \hii regions
(e.g., \citealt{deharveng2010,jose2013,samal2014,deharveng2015}). In Fig. \ref{spatial}, we find that the WISE W4 band (22 $\mu$m) emission
from warm dust is strongest in the vicinity of OB stars associated with the cluster Stock 8.  Similarly, in the CGPS catalog of extended
radio sources \citep{kerton2007} as well as in the NVSS  1.4 GHz survey \citep{condon1998}, only Stock 8 is found to be associated with the
radio continuum free-free emission (see Fig. \ref{spatial}).  These results suggest that the \hii region (Sh2-234) associated with Stock 8
is likely younger than the surrounding massive stars.

\subsection{Radiation field around Stock 8}
 Stock 8 is located in the  harsh environment of an  expanding bubble and  pool of older
  OB stars which may have a considerable effect on the structure of the  molecular material  around it.
  We measure the total  EUV flux reaching at  the center of Stock 8 from the Lyman continuum photons emitted per second  
  from  the massive stars  around the cluster. We obtained the number of ionizing photons
  for  O8-B3V spectral types from   \cite{martins2005} and \cite{panagia1973}. 
  The total flux is estimated using the projected distance between each massive star to the cluster center
  and is $\sim$ 5 $\times$ 10$^{11}$   photons/s/cm$^{-2}$. Here we ignore the  possible attenuation by diffuse gas and blocking by molecular clumps
  and also the uncertainty due to the use of projected distance. The total FUV flux  reaching at the center of Stock 8 is
  measured in terms of the Habing flux G0 (equal to 1.6 $\times$ 10$^{-3}$ erg cm$^{-2}$s$^{-1}$)   and  is   3 $\times$ 10$^4$ $G_0$. 
    The FUV strength is much higher than   the average  interstellar radiation field (i.e.,  1.7 $\times$ $G{_0}$; \citealt{habing1968}).
  
 The kinetic energy of the   expanding shell is $\sim$  3$\times$ 10$^{50}$ erg \citep{kang2012}.
 In the absence of any high resolution molecular observations towards the region, we use the low resolution 
 $^{12}$CO map by \cite{leisawitz1989} to roughly  measure the internal kinetic energy ($E_{kin}$, 1/2 $M_{cloud}$ $\sigma^2$) 
 of the molecular cloud  associated with Stock 8 and nebular region (mass  $\sim$ 3.4 $\times$ 10$^3$ M$_\odot$;
 resolution $\sim$ 8$^\prime$.7; \citealt{leisawitz1989}), where
  $\sigma$ represents the turbulence due to internal random motions in the molecular cloud.  $\sigma$ is calculated using the
  Larson's relations \citep{larson1981}, for the  effective radius of the cloud  $\sim$ 20 pc.
  The $E_{kin}$ of the cloud is estimated to be $\sim$ 4 $\times$ 10$^{47}$ ergs, which is lower than the kinetic  energy of the expanding shell.
   
  The above analysis shows that the external environment might be expected to have an impact on the structure  of the molecular cloud
  around Stock 8. However, our analysis on the low mass regime  of the IMF of Stock 8 shows  that  there is no strong evidence for
a difference in the underlying IMF  between Stock 8 and other resolved star forming regions and the
Galactic field. These results likely suggest that   the role of external feedback has only a weak role in the shape of IMF and star
formation process in Stock 8.

\section{Summary}

Stock 8 is a young cluster associated with the \hii region Sh2-234, and located at the edge of an expanding  giant shell. In this study,
we analysed the stellar content and star formation history of the cluster  using a  sensitive $V$, $I$-band  ($V$ $\sim$ 24 mag) optical
photometry, complemented with the UKIDSS-$JHK$ and 3.6, 4.5 $\mu$m {\it Spitzer}-IRAC  photometry. The optical CMD extends to substantially
fainter magnitudes than previous studies, sampling much of the main sequence  as well as pre-main sequence  objects and the NIR CMD extends
to  masses as low as $\sim$ 0.08 $M_{\odot}$. Using near and mid-infrared photometry, we obtain the census  of the young stellar  population and their
characteristics within 3$^\prime$ radius of the  cluster, which  includes 187  candidate Class I and Class II YSOs. The fraction of sources
having circumstellar disk is $\sim$ 35\% in the mass range  1.5-0.4 $M_{\odot}$. Both optical and NIR color-magnitude diagrams show a well
defined sequence of young stellar objects. From the optical color-magnitude diagram analysis using evolutionary models of various ages by \citet{siess2000},
we obtain a median log(age) of $\sim$ 6.5 ($\sim$  3.0 Myr) with an age spread of $\sim$ 0.25 dex for the cluster. Through Monte Carlo simulations
we quantify the effective broadening in the synthetic color-magnitude diagrams ($\sim$ 0.15 dex) by various factors which causes the age spread.
 The  intrinsic  age spread  ($\sim$ 0.2 dex) evident from the color-magnitude diagram analysis is comparable to the other star
  forming regions. 

The $K$-band luminosity function of Stock 8 after field star decontamination and completeness correction shows a broad peak between $\sim$ 14.5
and 16.5 mag and  steadily declines to  fainter, less-massive objects. The $K$-band  luminosity function of Stock 8 is comparable to that of the
Trapezium cluster. IMF of Stock 8 is well described by a log-normal distribution with a  characteristic mass ($m_0$=0.4M$_\odot$), which is
comparable with those found in other well studied nearby regions. Above 0.5M$_\odot$, the IMF is well consistent  with the Salpeter slope.  Though
Stock 8 is surrounded by several massive stars, we conclude that there seems to be no severe environmental effect in the form of IMF due to the
proximity of the massive stars. 

\acknowledgments
We are grateful to the anonymous referee for the useful  comments which have helped us to improve the scientific contents of the paper.
This paper is based on data obtained as part of the UKIRT Infrared Deep Sky Survey.  This publication made use of the data products from
SIMBAD database (operated at CDS, Strasbourg, France), Two Micron All Sky Survey (a joint project of the University
of Massachusetts and the Infrared Processing and Analysis Center/California Institute of Technology, funded by NASA and NSF), 
archival data obtained with the {\it Spitzer} Space Telescope (operated by the Jet Propulsion Laboratory, California Institute of Technology 
under a contract with NASA), {\it Wide-Field Infrared Survey Explorer} (a joint project of the University of California, Los Angeles, 
and the Jet Propulsion Laboratory (JPL), California Institute of Technology (Caltech), funded by the National Aeronautics and Space
Administration (NASA)) and NOAO Science archive, which is operated by the Association of Universities for Research in Astronomy (AURA),
Inc. under a cooperative agreement with the National Science Foundation.
This work is supported by a Youth Qianren grant to G.J.H. and general grant \# 11473005 awarded by the National Science
Foundation of China. JJ acknowledges the General Financial Grant \# 2015M570883 and Special Financial Grant \# 2016T90008 from the China Postdoctoal 
Science Foundation.  JJ would like to thank A.K. Pandey for useful discussion in the beginning and Guo Zhen, Long Feng,  Yuguang Chen for
helping with the SNIFS observations.

%\bibliographystyle{apj}
%\bibliography{ref}
%\bsp

\clearpage
\end{document}